\documentclass[11pt]{article}

\usepackage[a4paper,margin=1in]{geometry}
\usepackage{amsmath,amssymb,mathtools,bm}
\usepackage{graphicx}
\usepackage{booktabs}
\usepackage{array}
\usepackage{multirow}
\usepackage{microtype}
\usepackage{enumitem}
\usepackage{cite}
\usepackage{xcolor}
\usepackage{hyperref}
\usepackage{cleveref}
\usepackage{siunitx}
\usepackage[normalem]{ulem} 
\usepackage{cancel}        

\newcommand{\Mpl}{\mathrm{M_{Pl}}}

\title{\bfseries Localized Neutrinos in Two Dark Dimensions}

\author{
Waqas Ahmed$^{1}$ \quad and\quad George K. Leontaris$^{2}$\\[1.5mm]
{\small $^{1}$Center for Fundamental Physics, School of Artificial Intelligence, Hubei Polytechnic University,}\\
{\small Huangshi 435003, China}\\[1mm]
{\small $^{2}$Physics Department, Theory Division, University of Ioannina,}\\
{\small 45110 Ioannina, Greece}
}

\date{}

\begin{document}

\maketitle
\begin{abstract}
We study neutrino masses in a six-dimensional framework with two compact
dark dimensions. In the conventional flat-bulk realization, the
sterile-neutrino zero mode is strongly volume suppressed, so a fundamental
scale near $10~\mathrm{TeV}$ requires a large higher-dimensional Yukawa
coupling to reproduce the observed neutrino masses, while the accompanying
two-dimensional Kaluza--Klein tower is strongly constrained by oscillation
data. We show that localizing the sterile-neutrino wave function near the
Standard-Model brane substantially alleviates both effects. Using a Gaussian localizer, realistic Dirac neutrino masses are obtained with perturbative, order-one Yukawa couplings, while the brane-visible sterile excitations form a shell spectrum with $M_s^2=8\mu^2 s$, and squared active--sterile mixings that decrease as $1/s$. The localization
scale therefore separates the experimentally relevant sterile-neutrino
spectrum from the gravitational compactification scale. We derive the
resulting three-flavour structure and study its phenomenological consequences for reactor oscillations and tritium beta decay, including comparisons with current flat-LED and KATRIN constraints.
\end{abstract}

\section{Introduction}

The origin of neutrino mass remains one of the central open questions beyond the Standard Model (SM). A simple possibility is that right-handed neutrinos are not confined to the four-dimensional brane on which the SM fields reside. Being singlets under the SM gauge group, they may propagate in extra spatial dimensions\cite{Dienes1999,ArkaniHamed2001}. Their wave functions are then normalized over the compact bulk, which can suppress the effective four-dimensional Yukawa interaction and naturally generate small Dirac neutrino masses
\cite{Dienes1999,ArkaniHamed2001,DvaliSmirnov1999,Lukas2001}.
This basic idea has been developed in several forms, including bulk-seesaw, unified, and interval constructions
\cite{KimRaby2003,EiseleHaba2006,WatanabeYoshioka2011,DiegoQuiros2008,Blennow2010}.
A characteristic consequence is a tower of Kaluza--Klein (KK) neutrino states. These states mix with the active neutrinos and make the extra-dimensional origin of neutrino mass experimentally testable.

The subject has received renewed attention in the context of the Dark Dimension proposal. In this framework, swampland-motivated arguments connect the observed cosmological constant with the existence of a mesoscopic compact dimension
\cite{MonteroVafa2023}.
More generally, micron- and sub-millimetre extra dimensions have a long history in TeV-scale gravity
\cite{Antoniadis1990,ADD1998,AntoniadisADD1998}
and are directly constrained by short-distance tests of Newtonian gravity
\cite{Lee2020,Murata2026}.
Neutrinos are especially sensitive probes of such scenarios because gauge-singlet fermions can explore the bulk without introducing SM gauge interactions.

A substantial number of recent studies have therefore investigated right-handed neeutrinos in large or dark dimensions using reactor, accelerator, long-baseline, and beta-decay experiments
\cite{Machado2011,Machado2012,Esmaili2014,GirardiMeloni2014,Stenico2018,BastoJUNO2022,Forero2022,Eller2025,DeGiorgiTurner2026,AntoniadisKATRIN2026,Bai2026,Franklin2026}.
Related constructions have connected the neutrino sector with $B-L$ dynamics, non-standard boundary conditions, and more general sterile towers
\cite{MonteroBminusL2026,Abe2026,Lonardi2026}.
A recent five-dimensional Majorana construction has also been studied in detail using Daya Bay, KATRIN, and KamLAND-Zen data
\cite{DeGiorgiPasari2026}.
That analysis illustrates an important point: the dimensionality and Lorentz structure of the bulk theory can qualitatively modify both the KK spectrum and its phenomenology.

Very recently, the cosmological consequences of bulk Majorana neutrinos in
the dark dimension were analyzed in Ref.~\cite{AnchordoquiVafa2026}.  In
that five-dimensional construction, thermal production of the sterile KK
tower can lead to the cosmological bulk-neutrino catastrophe unless the
tower is sufficiently decoupled; for the particular thermal history and
Majorana setup considered there, a bulk Majorana mass of order
\(100~\mathrm{GeV}\) suppresses the unwanted population.  This numerical
scale is not a bound on the present six-dimensional Dirac construction,
whose spectrum and density of brane-visible states are different.  It does,
however, make the early-Universe production of the localized 2DD shells an
important consistency test of the model.


In the present work, we explore a higher-dimensional extension of the Dark Dimension scenario~\cite{Anchordoqui:2025nmb}. Specifically, we consider the two-dark-dimensions (2DD) framework studied in Ref.~\cite{AhmedLeontaris2026}, in which gravity propagates in six spacetime dimensions and the fundamental gravitational scale can lie near $\mathcal{O}$(10)\,TeV.
 Two compact dimensions change the neutrino problem in an essential way. For a flat bulk fermion, the sterile zero mode is diluted over the two-dimensional compact volume. Its value on the SM brane is therefore small, leading to a strong suppression of the effective Dirac mass. At the same time, the number of KK states grows with energy because the modes populate a two-dimensional lattice. Codimension-two brane interactions also display logarithmic ultraviolet sensitivity
\cite{Dudas2005}.
These features make the flat 2DD neutrino sector considerably more restrictive than its simplest one-dimensional counterpart.

There is then a simple tension. Increasing the higher-dimensional Yukawa
coupling can compensate for the small zero-mode overlap and reproduce the
observed neutrino masses. However, the same interaction couples the active
neutrinos to the KK tower. A larger Yukawa coupling therefore also enhances
the sterile admixture. This is potentially problematic because flat
extra-dimensional neutrino models are already tightly constrained by
disappearance measurements
\cite{MINOS2019,DayaBaySterile2024,Forero2022}.
For two equal flat extra dimensions, recent reactor analyses find, for a small lightest neutrino mass,
\[
R<0.45~\mu\mathrm{m}\qquad (\mathrm{NO}),
\]
and
\[
R<0.09~\mu\mathrm{m}\qquad (\mathrm{IO}),
\]
at 99\% C.L.
\cite{Elacmaz2025}.
These limits apply to flat bulk wave functions. They should not be directly transferred to a localized sterile sector. They nevertheless make clear why the neutrino sector requires additional structure if the gravitational radius is to remain in the micron-scale region.

The possibility explored here is localization. Fermion localization in extra dimensions has long been used to generate effective Yukawa hierarchies from wave-function overlaps
\cite{ArkaniHamedSchmaltz2000,MirabelliSchmaltz2000,Girmohanta2021}.
We apply the same idea to sterile neutrinos in two dark dimensions. Instead of allowing the right-handed neutrino zero mode to spread uniformly over the compact space, we localize it near the SM brane. A Gaussian profile provides a simple analytic realization and can be viewed as the local approximation to trapping by a smooth scalar or vortex background
\cite{Girmohanta2021}.
 
This modification has two important consequences. First, localization increases the value of the normalized zero-mode wave function on the brane. The effective Dirac mass is therefore enhanced, and realistic neutrino masses can be obtained without taking the six-dimensional Yukawa coupling to be large. Second, the sterile excitation spectrum is no longer set directly by $R^{-1}$. It is controlled instead by the localization scale $\mu$. The gravitational compactification radius and the neutrino excitation scale therefore become physically distinct quantities.

The localized spectrum also has a particularly simple structure. In the harmonic approximation, the transverse excitations form a two-dimensional oscillator spectrum. Many states are degenerate, but a pointlike brane couples only to specific even-parity combinations. After the degenerate states are rotated to the brane-coupled basis, only one visible combination appears at each shell. This reduces the apparently complicated two-dimensional tower to a simple sequence of sterile states. Their masses and active--sterile mixings can be obtained analytically. The total sterile admixture then grows only logarithmically with the number of shells below the validity scale of the localized effective theory.

The main purpose of this work is therefore not simply to introduce another KK neutrino spectrum. Rather, we ask whether localization can separate the neutrino sector from the gravitational compactification scale while retaining the characteristic two-dark-dimension geometry. We show that this can be achieved within a simple Gaussian benchmark. For a fundamental scale near $10~\mathrm{TeV}$, the observed neutrino masses can be reproduced with perturbative Yukawa couplings, while the first brane-visible sterile states are shifted to masses controlled by $\mu$. The resulting framework leads to distinctive signatures in oscillation experiments and in tritium beta decay.

The remainder of the paper is organized as follows. In Sec.~\ref{sec:geometry} we introduce the two-dark-dimensions geometry and quantify the neutrino-mass and KK-mixing problem of the flat-bulk realization. In Sec.~\ref{sec:localizer} we construct the localized sterile-neutrino sector and derive the enhancement of the effective Dirac mass. The localized excitation spectrum and its brane overlaps are obtained in Sec.~\ref{sec:spectrum}, and the three-flavour embedding is discussed in Sec.~\ref{sec:flavour}. Reactor-oscillation phenomenology is presented in Sec.~\ref{sec:osc}, while the beta-decay signatures are studied in Sec.~\ref{sec:beta}. We comment on Majorana neutrinos and lepton-number violation in Sec.~\ref{sec:majorana}, and collect the representative numerical results in Sec.~\ref{sec:numerics}. We conclude in the final section.

\section{Two Dark Dimensions and the flat-neutrino problem}
\label{sec:geometry}

We consider a six-dimensional spacetime in which the two extra dimensions
form a compact space,
\begin{equation}
\mathcal M_6=\mathcal M_4\times\mathcal C_2,
\end{equation}
where \(\mathcal C_2\) is a compact two-dimensional manifold.  For explicit
numerical estimates we take \(\mathcal C_2\) to be an equal-radius square
torus, or its orbifold descendant, so that its volume is
\begin{equation}
V_2=(2\pi R)^2.
\label{eq:V2}
\end{equation}
The four-dimensional Planck mass \(\Mpl\) is then related to the
six-dimensional gravity scale \(M_6\) by the standard dimensional-reduction
relation
\begin{equation}
\Mpl^2=M_6^4V_2.
\label{eq:gravityrelation}
\end{equation}
For equal radii this fixes the compactification radius in terms of the two
mass scales,
\begin{equation}
R=\frac{\Mpl}{2\pi M_6^2},
\qquad
R^{-1}=\frac{2\pi M_6^2}{\Mpl}.
\label{eq:Rrelation}
\end{equation}
Taking \(\Mpl=2.435\times10^{18}~\mathrm{GeV}\) and \(M_6=10~\mathrm{TeV}\)
gives the benchmark values
\begin{equation}
R\simeq0.765~\mu\mathrm m,
\qquad
R^{-1}\simeq0.258~\mathrm{eV}.
\label{eq:Rbenchmark}
\end{equation}
The precise numerical coefficient depends on the compactification convention,
but the parametric relation \(R\propto M_6^{-2}\) is robust. The two-dark-dimension (2DD) scenario of Ref.~\cite{AhmedLeontaris2026} is
motivated by a fundamental scale of order \(10~\mathrm{TeV}\).  In the
gravitational sector, the compactification radius controls the transition
between four- and six-dimensional behavior.  The neutrino sector, however,
need not have a wave function that is flat over the full compact space.  As
we will see, this distinction is central to the viability of the model.

\subsection{The flat-zero-mode problem}

Let \(\Psi_i(x,y_1,y_2)\) be gauge-singlet six-dimensional fermions, and
suppose that the Standard Model (SM) lepton doublets \(L_\alpha\) and the
Higgs field \(H\) are localized at the point \(y_1=y_2=0\) of the compact
space.  A brane-localized Yukawa interaction can then be written
schematically as
\begin{equation}
S_Y=-\int d^4x\,d^2y\;\delta^{(2)}(\bm y)
\left[
\frac{(Y_6)_{\alpha i}}{M_6}
\overline L_\alpha\widetilde H\,\Psi_{R,i}(x,\bm y)
+\mathrm{h.c.}
\right].
\label{eq:6Dyukawa}
\end{equation}
The factor \(M_6^{-1}\) follows from dimensional analysis: in six dimensions
the fermion has mass dimension \(5/2\), while the brane fields retain their
four-dimensional dimensions. The coupling of the bulk fermion to the brane is controlled by the value of
its zero-mode wave function at the brane.  Let us denote the normalized zero
mode by \(\chi_0(\bm y)\).  For a flat zero mode,
\begin{equation}
\chi_0^{\rm flat}(\bm y)=\frac{1}{\sqrt{V_2}},
\end{equation}
so that after electroweak symmetry breaking achieved by the Higgs vev, \(\langle H\rangle=v/\sqrt2\),
the four-dimensional Dirac mass matrix reads
\begin{equation}
(m_D^{\rm flat})_{\alpha i}
=\frac{(Y_6)_{\alpha i}v}{\sqrt2M_6\sqrt{V_2}}.
\label{eq:mDflat1}
\end{equation}
Using Eq.~\eqref{eq:gravityrelation} to eliminate \(V_2\), this becomes
\begin{equation}
(m_D^{\rm flat})_{\alpha i}
=\frac{(Y_6)_{\alpha i}v}{\sqrt2}
\frac{M_6}{\Mpl}.
\label{eq:mDflat}
\end{equation}
Numerically,
\begin{equation}
 m_D^{\rm flat}
 \simeq7.14\times10^{-4}\,y_6
 \left(\frac{M_6}{10~\mathrm{TeV}}\right)\mathrm{eV},
\label{eq:mDflatnum}
\end{equation}
where \(y_6\) denotes a typical entry of the dimensionless coupling matrix
\((Y_6)_{\alpha i}\).  Consequently, reproducing a Dirac mass of
\(0.05~\mathrm{eV}\) at \(M_6=10~\mathrm{TeV}\) requires
\begin{equation}
 y_6\simeq70.
\label{eq:largey}
\end{equation}
Thus, the first problem of the flat 2DD realization is that the flat zero mode is strongly volume-suppressed,  requiring one to resort to a non-perturbatively large Yukawa coupling.

\begin{figure}[t]
\centering
\includegraphics[width=0.82\textwidth]{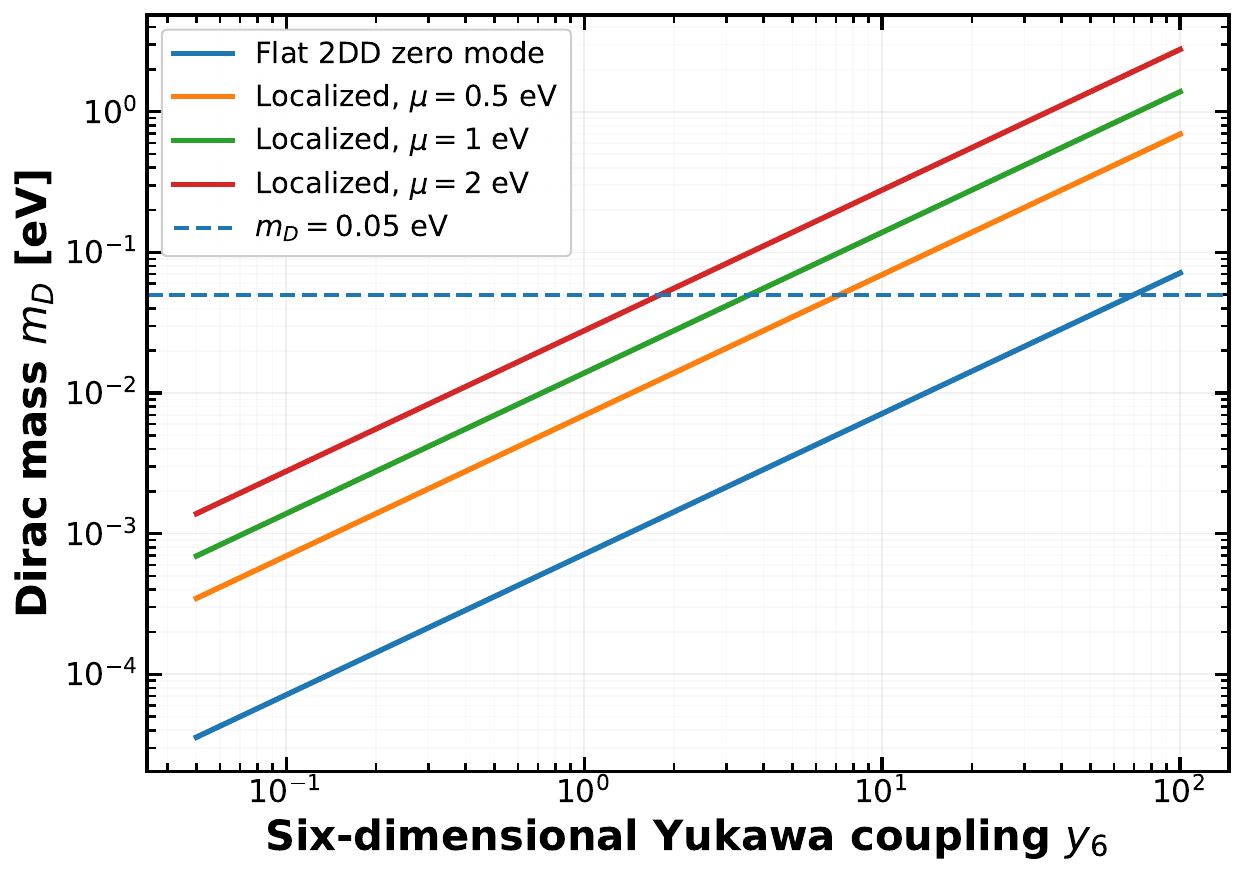}
\caption{Dirac mass as a function of the six-dimensional Yukawa coupling
\(y_6\) for \(M_6=10~\mathrm{TeV}\).  The flat 2DD zero mode (lower curve) is
strongly volume suppressed, requiring \(y_6\simeq70\) to reach
\(m_D\sim0.05~\mathrm{eV}\).  Localizing the sterile zero mode with scale
\(\mu\) (upper curve) increases its value at the brane and permits
\(m_D\sim0.05~\mathrm{eV}\) with perturbative couplings.}
\label{fig:mdy}
\end{figure}

The result is displayed in Fig.~\ref{fig:mdy}.  Raising \(M_6\) would
increase \(m_D\), but Eq.~\eqref{eq:Rrelation} then decreases the
compactification radius and weakens the connection to the micron-scale 2DD
regime.  For example, \(y_6\simeq1\) and \(m_D\simeq0.05~\mathrm{eV}\) would
require \(M_6\) in the several-hundred-TeV range.  As we discuss below,
localizing the sterile zero mode near the brane removes this suppression and
restores perturbative couplings.

\subsection{Kaluza--Klein tower and reactor bounds}

A flat bulk neutrino contains a two-dimensional Kaluza--Klein (KK) lattice
with masses
\begin{equation}
M_{n_1n_2}^2=\frac{n_1^2+n_2^2}{R^2}.
\end{equation}
The number of modes below energy \(E\) grows approximately as
\begin{equation}
N_{\rm KK}^{\rm flat}(E)\simeq\pi E^2R^2,
\label{eq:flatcount}
\end{equation}
and the increasing degeneracy enhances active--sterile oscillation effects.
The most recent reactor analysis including Daya Bay, RENO, KamLAND, NEOS and
STEREO derived, for equal radii and a small lightest neutrino mass, the
\(99\%\) C.L. limits \cite{Elacmaz2025}
\begin{equation}
R<0.45~\mu\mathrm m\quad(\mathrm{NO}),
\qquad
R<0.09~\mu\mathrm m\quad(\mathrm{IO}),
\label{eq:flatbounds}
\end{equation}
where NO and IO denote normal and inverted neutrino mass ordering,
respectively.  The same work found that the \(d=2\) reactor sensitivity is
dominated by Daya Bay, with a smaller contribution from RENO.  For
comparison, Ref.~\cite{Forero2022} combined reactor, MINOS/MINOS+, Daya Bay
and KATRIN data in a different model with one dominant flat extra dimension
and obtained \(R_{\rm ED}<0.20~\mu\mathrm m\) (NO) and
\(R_{\rm ED}<0.10~\mu\mathrm m\) (IO) at \(90\%\) C.L.  These results are
useful benchmarks for the experimental scale, but they are not bounds on the
localized 2DD construction.

\begin{figure}[t]
\centering
\includegraphics[width=0.82\textwidth]{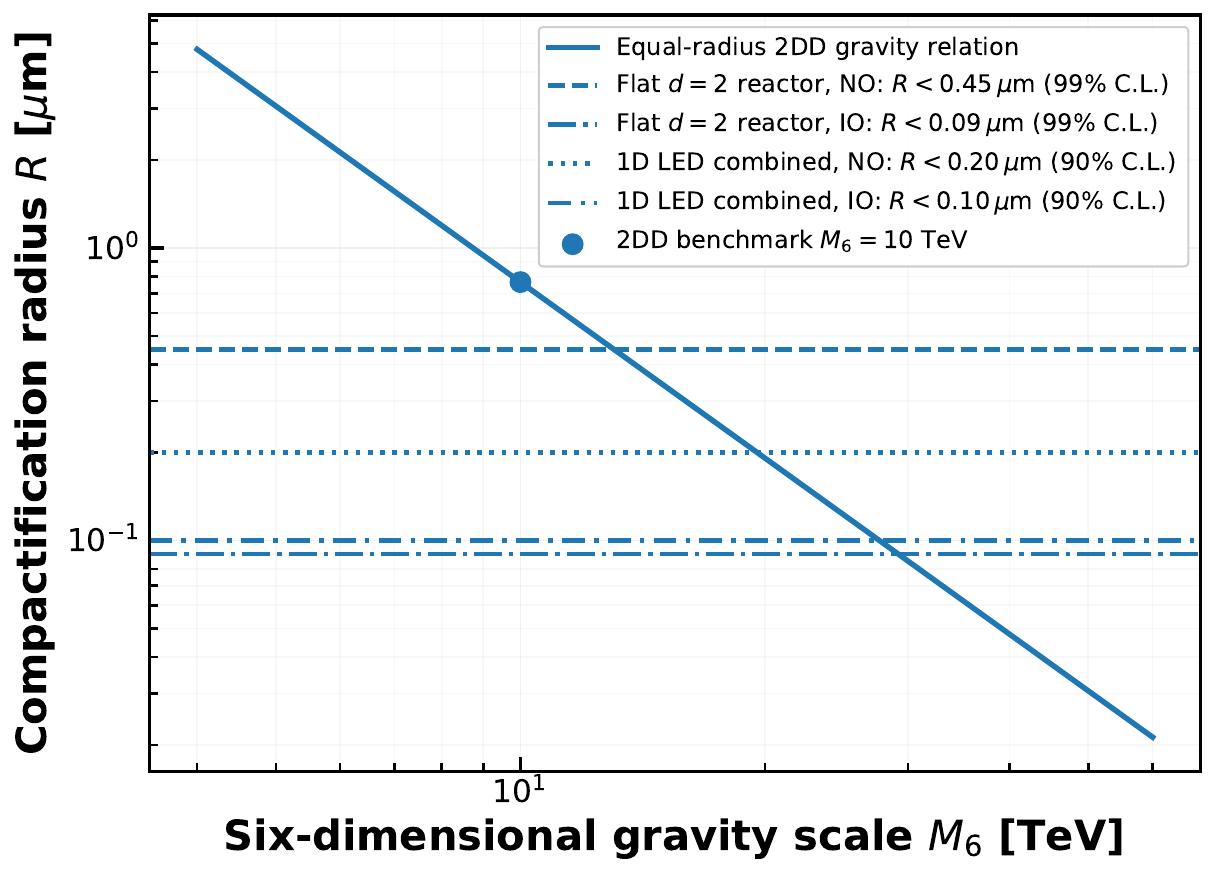}
\caption{Published flat-LED radius bounds shown against the equal-radius 2DD
gravity relation of Eq.~\eqref{eq:Rrelation}.  The \(d=2\) reactor limits
are from Ref.~\cite{Elacmaz2025}; the one-dominant-dimension combined limits
are from Ref.~\cite{Forero2022}.  Because both analyses assume flat bulk wave
functions, the lines are comparison benchmarks rather than exclusions of the
localized model.  The \(M_6=10~\mathrm{TeV}\) benchmark of
Eq.~\eqref{eq:Rbenchmark} lies below the flat-neutrino region, illustrating
the tension that motivates localization.}
\label{fig:rbounds}
\end{figure}

Under the equal-radius gravity relation, Eq.~\eqref{eq:flatbounds}
corresponds roughly to
\begin{equation}
M_6\gtrsim13.0~\mathrm{TeV}\quad(\mathrm{NO}),
\qquad
M_6\gtrsim29.1~\mathrm{TeV}\quad(\mathrm{IO}),
\label{eq:M6flatbound}
\end{equation}
if the neutrino is assumed to be a flat bulk field.  The \(M_6=10~\mathrm{TeV}\)
point of Eq.~\eqref{eq:Rbenchmark} therefore lies outside the flat-neutrino
benchmark region.  This comparison is shown in Fig.~\ref{fig:rbounds}.  These
are not bounds on the localized model developed below; rather, they quantify
the tension that localization is designed to remove.

\section{Localized sterile neutrinos}
\label{sec:localizer}

Fermion localization provides a direct way to change the value of a
normalized bulk wave function at the SM brane.  Localized profiles in extra
dimensions have long been used to generate hierarchical effective Yukawa
couplings from order-one higher-dimensional parameters
\cite{ArkaniHamedSchmaltz2000,MirabelliSchmaltz2000}.

Different microscopic trapping mechanisms need not produce identical
excitation spectra.  We therefore do not identify the Gaussian profile with
a unique ultraviolet completion.  Instead, we use the fact that near a smooth,
non-degenerate minimum of an approximately isotropic trapping potential the
low-lying transverse dynamics can be described by a two-dimensional harmonic
oscillator.  Gaussian localized fermion profiles in two extra dimensions
provide a standard calculational benchmark for such a low-energy description
\cite{Girmohanta2021}.
To implement the harmonic localizer consistently, define
\begin{equation}
\mathcal D_a=\partial_a+2\mu^2y_a,
\qquad a=1,2,
\label{eq:Doperator}
\end{equation}
so that
\begin{equation}
\mathcal H_\perp
=\sum_{a=1}^{2}\mathcal D_a^\dagger\mathcal D_a
=-\nabla_\perp^2+4\mu^4r^2-4\mu^2,
\qquad
r^2=y_1^2+y_2^2.
\label{eq:Hperp}
\end{equation}
The operator is positive semidefinite by construction.  Its lowest mode
satisfies \(\mathcal D_a\chi_{00}=0\), and hence has zero transverse
eigenvalue before electroweak symmetry breaking.  The parameter \(\mu\)
sets the inverse localization length: increasing \(\mu\) narrows the
wave function around the brane.
We write the sterile field as
\begin{equation}
\Psi_{R,i}(x,\bm y)
=\sum_A N_{R,i}^{(A)}(x)\,\chi_A(\bm y),
\label{eq:localexp}
\end{equation}
with
\begin{equation}
\int_{\mathcal C_2}d^2y\;\chi_A^*(\bm y)\chi_B(\bm y)=\delta_{AB}.
\end{equation}
The four-dimensional brane coupling to mode \(A\) is then
\begin{equation}
(m_D^{A})_{\alpha i}
=\frac{(Y_6)_{\alpha i}v}{\sqrt2M_6}\chi_A(0).
\label{eq:modecouplinggeneral}
\end{equation}
The problem is therefore reduced to determining the localized spectrum and
the brane overlaps \(\chi_A(0)\).
This construction is an effective low-energy description rather than a
complete six-dimensional ultraviolet model.  A microscopic completion must
specify the six-dimensional spinor content and the background responsible
for localization.  The laboratory results below depend only on the normalized
low-lying wave functions and are therefore insensitive to those ultraviolet
details as long as the harmonic approximation is valid for the states being
probed.

In particular, $\mathcal H_\perp$ should be understood as the effective
squared transverse operator governing the brane-coupled fermionic component,
rather than as a complete microscopic six-dimensional fermion Lagrangian.
The zero mode satisfies
\begin{equation}
\mathcal D_a\chi_{00}=0,
\end{equation}
so that
\begin{equation}
\chi_{00}(r)=A_0\exp(-\mu^2r^2).
\label{eq:gaussian}
\end{equation}
Normalization on the compact square gives
\begin{align}
1
&=A_0^2
\left[\int_{-\pi R}^{\pi R}dy\;e^{-2\mu^2y^2}\right]^2
=A_0^2
\left[
\frac{\sqrt{\pi}}{\sqrt{2}\mu}\,
\operatorname{erf}(\sqrt{2}\pi\mu R)
\right]^2 .
\end{align}
which gives
\begin{equation}
A_0=
\sqrt{\frac{2}{\pi}}\,
\frac{\mu}{\operatorname{erf}(\sqrt2\pi\mu R)}.
\label{eq:Anorm}
\end{equation}
When $\mu R\gtrsim1$, the error function is already close to unity, $ \operatorname{erf}(x)\approx 1-\frac{e^{-x^2}}{\sqrt{\pi} x}$, and
\begin{equation}
\chi_{00}(0)\simeq\sqrt{\frac{2}{\pi}}\,\mu.
\label{eq:chi00}
\end{equation}
Strictly speaking, a wave function on a torus should be periodic.  The compact-space profile can be represented by the image sum
\begin{equation}
\chi_{00}^{\rm per}(\bm y)\propto
\sum_{\bm \ell\in\mathbb Z^2}
\exp\!\left[-\mu^2\left|\bm y+2\pi R\bm\ell\right|^2\right].
\label{eq:periodicGaussian}
\end{equation}
For $\mu R\gg1$ the images are exponentially suppressed near the brane, so the single Gaussian used in the analytic derivation is the leading term of Eq.~\eqref{eq:periodicGaussian}.  We also assume that the SM brane is centered on the localized profile; displacing it would remove the simple odd-mode selection rule used below.  For the benchmark used below, $\mu R\simeq4.65$, so finite-volume image corrections are negligible for the low-lying modes.
Substituting Eq.~\eqref{eq:chi00} into Eq.~\eqref{eq:modecouplinggeneral}, the localized zero-mode Dirac matrix is
\begin{equation}
(m_D^{\rm loc})_{\alpha i}
=\frac{(Y_6)_{\alpha i}v}{\sqrt\pi M_6}\,\mu.
\label{eq:mDloc}
\end{equation}
The ratio to the flat result is
\begin{equation}
\xi_{\rm loc}
\equiv
\frac{m_D^{\rm loc}}{m_D^{\rm flat}}
\simeq
2\sqrt{2\pi}\,\mu R,
\label{eq:enhancement}
\end{equation}
For finite $\mu R$ the exact corresponding ratio within the Gaussian approximation is
\begin{equation}
\xi_{\rm loc}^{\rm exact}=
\frac{2\sqrt{2\pi}\,\mu R}
{\operatorname{erf}(\sqrt2\pi\mu R)},
\end{equation}
which rapidly approaches Eq.~\eqref{eq:enhancement} once the profile is well localized. Thus localization provides the enhancement needed to avoid the large-Yukawa
problem of the flat zero mode.  Equation~\eqref{eq:Rrelation} is needed only
when the result is subsequently rewritten in terms of
\(\kappa_{\rm loc}=\mu R\) and \(M_6/\Mpl\). It is useful to trade off the apparently small localization scale for the dimensionless ratio
\begin{equation}
\kappa_{\rm loc}\equiv\mu R.
\end{equation}
If the trapping dynamics is tied to compactification, $\mu=\kappa_{\rm loc}/R$, then Eq.~\eqref{eq:Rrelation} gives
\begin{equation}
m_D^{\rm loc}=2\sqrt\pi\,\kappa_{\rm loc}\,y_6\,v\frac{M_6}{\Mpl}.
\label{eq:mDkappa}
\end{equation}
The benchmark $\mu=1.2~\mathrm{eV}$ at $M_6=10~\mathrm{TeV}$ corresponds to $\kappa_{\rm loc}\simeq4.65$. In this interpretation, the model does not introduce an unrelated eV-scale input; rather, it requires a localization width that is smaller than the compactification radius by a factor of a few. A microscopic construction should ultimately explain why $\kappa_{\rm loc}$ takes a value of this order. Nevertheless, this requirement concerns a dimensionless geometric ratio rather than a new hierarchy of order $\mu/M_6$. For $M_6=10~\mathrm{TeV}$, this corresponds to a localization width of approximately $0.22$ times the compactification radius.

\begin{equation}
m_D^{\rm loc}
\simeq
1.388\times10^{-2}\,y_6
\left(\frac{\mu}{\mathrm{eV}}\right)\mathrm{eV}.
\label{eq:mDlocnum}
\end{equation}
The atmospheric scale can therefore be obtained with
\begin{equation}
y_6\simeq3.60
\left(\frac{m_D}{0.05~\mathrm{eV}}\right)
\left(\frac{\mathrm{eV}}{\mu}\right)
\left(\frac{M_6}{10~\mathrm{TeV}}\right).
\label{eq:yreq}
\end{equation}
A localizer scale near $1~\mathrm{eV}$ already brings the largest Yukawa eigenvalue into the perturbative range while remaining larger than the compactification scale $R^{-1}\simeq0.258~\mathrm{eV}$.  For fixed $M_6$, the required Yukawa eigenvalue simply scales as $y_i\propto m_i/\mu$, so the improvement does not rely on an additional numerical tuning.


\section{Localized excitation spectrum and brane overlaps}
\label{sec:spectrum}
We instead use the positive operator already defined in
Eq.~\eqref{eq:Hperp}.  Its eigenfunctions factorize,
\begin{equation}
\chi_{n_1n_2}(y_1,y_2)
=
\varphi_{n_1}(y_1)\varphi_{n_2}(y_2),
\end{equation}
where
\begin{equation}
\varphi_n(y)=
\left(\frac{\beta}{\pi}\right)^{1/4}
\frac{H_n(\sqrt{\beta}\,y)}{\sqrt{2^n n!}}\,
e^{-\beta y^2/2},
\qquad
\beta=2\mu^2.
\label{eq:oscwf}
\end{equation}
For the one-dimensional operator
\(-\partial_y^2+4\mu^4y^2-2\mu^2\), the eigenvalue is
\(4\mu^2n\).  Therefore the two-dimensional transverse masses are
\begin{equation}
M_{n_1n_2}^2
=
4\mu^2(n_1+n_2),
\qquad
n_1,n_2=0,1,2,\ldots .
\label{eq:localmasses}
\end{equation}
The \((0,0)\) mode has zero transverse mass before electroweak symmetry
breaking and has the Gaussian profile already given in
Eq.~\eqref{eq:gaussian}. The lightest transverse excitations instead satisfy
\begin{equation}
M_{1,0}^2=M_{0,1}^2=4\mu^2,
\qquad
M_{1,0}=M_{0,1}=2\mu .
\end{equation}
As shown below, these first excited states are odd in one transverse
direction and vanish at a pointlike brane located at the origin.
Because the Hermite functions have definite parity,
\(\varphi_n(-y)=(-1)^n\varphi_n(y)\), the brane overlaps satisfy
\begin{equation}
\chi_{n_1n_2}(0)=
\varphi_{n_1}(0)\,\varphi_{n_2}(0),
\qquad
\varphi_n(0)=0 \ \text{for odd } n .
\end{equation}

More generally, every state for which either \(n_1\) or \(n_2\) is odd
vanishes at the origin.  Only even--even states have a nonzero pointlike
brane overlap,
\[
\varphi_{2k}(0)\propto
\left(\frac{\beta}{\pi}\right)^{1/4}
\frac{H_{2k}(0)}{\sqrt{2^{2k}(2k)!}} .
\]
This parity selection rule removes the \(n_1+n_2=1\) level from the
brane-coupled spectrum.
At the origin,
\begin{equation}
H_{2k+1}(0)=0,
\qquad
H_{2k}(0)=(-1)^k\frac{(2k)!}{k!},
\end{equation}
so every state with an odd quantum number in either direction vanishes at the brane.  Defining the overlap ratio
\begin{equation}
c_{n_1n_2}\equiv
\frac{\chi_{n_1n_2}(0)}{\chi_{00}(0)},
\end{equation}
we obtain
\begin{equation}
c_{2k,2\ell}
=(-1)^{k+\ell}
\frac{\sqrt{(2k)!}}{2^k k!}
\frac{\sqrt{(2\ell)!}}{2^\ell \ell!},
\label{eq:coverlap}
\end{equation}
with $c_{n_1n_2}=0$ if either index is odd.  The squared one-dimensional coefficient is
\begin{equation}
a_k\equiv
\frac{(2k)!}{4^k(k!)^2}
=\frac{1}{4^k}\binom{2k}{k}.
\label{eq:ak}
\end{equation}

The first oscillator states at mass $2\mu$ are odd in one direction and do not couple to a pointlike brane.  The first brane-visible states are $(2,0)$ and $(0,2)$, with
\begin{equation}
M_{\rm vis,1}=2\sqrt2\,\mu.
\label{eq:firstvisible}
\end{equation}
This selection rule is phenomenologically useful because it raises the first directly coupled sterile threshold relative to the basic localization scale.
The localized tower simplifies once the degeneracy on each shell is treated
correctly.  A brane-visible state has $n_1=2k$ and $n_2=2\ell$, so define the
shell index
\begin{equation}
    s=k+\ell=1,2,\ldots .
\end{equation}
All states on shell $s$ have the same mass,
\begin{equation}
    M_s^2 = 4\mu^2(2s) = 8\mu^2 s,
    \label{eq:shellmass}
\end{equation}
because $n_1+n_2 = 2(k+\ell) = 2s$.
Their brane overlaps are not independent.  From
Eqs.~\eqref{eq:coverlap}--\eqref{eq:ak}, the squared overlap of the state
$(2k,\,2(s-k))$ factorizes,
\begin{equation}
    \bigl|c_{2k,\,2(s-k)}\bigr|^2
    =
    a_k\,a_{s-k},
    \qquad
    a_k \equiv \frac{(2k)!}{4^k(k!)^2}.
    \label{eq:factorizedoverlap}
\end{equation}
The sequence $a_k$ satisfies a convolution identity that can be derived from
its generating function.  Using the binomial series,
\begin{equation}
    g(x) \equiv \sum_{k=0}^{\infty} a_k x^k
    =
    \sum_{k=0}^{\infty}\binom{2k}{k}\left(\frac{x}{4}\right)^{k}
    =
    \frac{1}{\sqrt{1-x}} .
    \label{eq:genfun}
\end{equation}
Squaring $g$ and expanding,
\begin{equation}
    g(x)^2
    =
    \sum_{s=0}^{\infty}
    \Bigl(\sum_{k=0}^{s} a_k\,a_{s-k}\Bigr) x^{s}
    =
    \frac{1}{1-x}
    =
    \sum_{s=0}^{\infty} x^{s}.
\end{equation}
Matching coefficients of $x^s$ gives the convolution identity
\begin{equation}
    \sum_{k=0}^{s} a_k\,a_{s-k} = 1
    \qquad \text{for every } s\ge 0 .
    \label{eq:convolution}
\end{equation}

Equation~\eqref{eq:convolution} has a direct interpretation.  Let $v$ be the
vector of brane overlaps on shell $s$,
$v_k \equiv c_{2k,\,2(s-k)}$.  Its squared norm is
\begin{equation}
    \sum_{k=0}^{s} |v_k|^2
    =
    \sum_{k=0}^{s} a_k\,a_{s-k}
    =
    1 ,
\end{equation}
so $v$ is a unit vector in the $(s+1)$-dimensional shell space.  A unitary
rotation whose first row is $v^\dagger$ therefore maps the shell basis to a
new basis in which exactly one combination $N_R^{(s)}$ has a nonzero brane
overlap, equal to $\chi_{00}(0)$, while all orthogonal combinations decouple
from the brane Yukawa interaction.  The coupling of this single visible
combination is then
\begin{equation}
    \Bigl[\sum_{k=0}^{s}
    \bigl|m_D\,c_{2k,\,2(s-k)}\bigr|^2\Bigr]^{1/2}
    =
    m_D
    \Bigl[\sum_{k=0}^{s} a_k\,a_{s-k}\Bigr]^{1/2}
    =
    m_D .
    \label{eq:shellcoupling}
\end{equation}
More precisely, at the full oscillator level $n_1+n_2=2s$ the degeneracy is
$2s+1$.  It consists of $s+1$ even--even states and $s$ odd--odd states.
The odd--odd states vanish at the brane from the outset.  Rotating the
$s+1$ even--even subspace therefore yields one brane-coupled combination and
$s$ orthogonal even--even combinations.  Thus the full level contains one
brane-visible state and $2s$ brane-inert states, without changing the
brane-coupled shell phenomenology derived above.
This shell structure is the central phenomenological result.  The
complicated degeneracy of the two-dimensional oscillator reduces, from the
point of view of the brane, to one visible state per shell, with a
nonuniform mass sequence $M_s\propto\sqrt{s}$ and a shell-independent Dirac
insertion $m_D$.
The pointlike-brane approximation is sufficient for the low-lying
phenomenology considered here.  If the SM brane has a smooth transverse
profile \(b(\bm y)\), normalized as
\(\int d^2y\,b(\bm y)=1\), the coupling to a transverse mode is determined by
the smeared overlap
\begin{equation}
\widetilde{\chi}_A
=
\int_{\mathcal C_2}d^2y\;
b(\bm y)\chi_A(\bm y).
\label{eq:smearedoverlap}
\end{equation}
For a brane width much smaller than the localization length,
\(\sigma_{\rm brane}\mu\ll1\), the first few overlaps approach their
pointlike values.  At sufficiently high excitation number a smooth brane
profile suppresses the overlap through transverse averaging.  Because that
ultraviolet suppression depends on the microscopic brane and localizer
structure, we do not introduce a separate brane-width form factor here.
Instead, its effect on the high-shell sum is encoded below in the effective
validity scale \(\Lambda_{\rm loc}\).

At the level of the four-dimensional effective theory we assume a vectorlike
six-dimensional Dirac completion.  For a suitable localizing background,
the nonzero transverse modes can be organized into paired sectors that form
massive four-dimensional Dirac states, whereas the existence and chirality
of the zero mode depend on the background and boundary conditions.  We choose
the latter such that the brane-coupled sterile zero mode has the
right-handed chirality required by the Yukawa interaction in
Eq.~\eqref{eq:6Dyukawa}.  The detailed six-dimensional spinor completion is
ultraviolet-model dependent and is not needed for the brane-overlap analysis.

Under this assumption, for a single generation, the Dirac mass matrix in the
basis where each shell has been rotated to its brane-coupled combination
takes the star form
\begin{equation}
\mathcal M_i=
\begin{pmatrix}
 m_{D,i} & m_{D,i} & m_{D,i} & m_{D,i} & \cdots\\
 0 & M_1 & 0 & 0 & \cdots\\
 0 & 0 & M_2 & 0 & \cdots\\
 0 & 0 & 0 & M_3 & \cdots\\
 \vdots & \vdots & \vdots & \vdots & \ddots
\end{pmatrix}.
\label{eq:starmatrix}
\end{equation}
The first column corresponds to the localized sterile zero mode, while the
remaining columns describe the visible shell combinations; the $s$ orthogonal even--even combinations and the additional
$s$ odd--odd states at the same full oscillator level do not appear because
they have zero brane coupling.  The physical Dirac masses are the singular
values of $\mathcal M_i$.
It is useful to make the perturbative limit explicit.  A finite truncation
to $N$ shells gives
\begin{equation}
\mathcal M_i\mathcal M_i^\dagger=
\begin{pmatrix}
 m_{D,i}^2(1+N) & m_{D,i}M_1 & m_{D,i}M_2 & \cdots\\
 m_{D,i}M_1 & M_1^2 & 0 & \cdots\\
 m_{D,i}M_2 & 0 & M_2^2 & \cdots\\
 \vdots & \vdots & \vdots & \ddots
\end{pmatrix},
\label{eq:MMdagger}
\end{equation}
where the first diagonal entry collects the zero-mode contribution
$m_{D,i}^2$ and the $N$ shell contributions $m_{D,i}^2$ from the repeated
first row.  For $m_{D,i}\ll M_s$, integrating out the heavy block gives
\begin{equation}
(m_i^{\rm phys})^2\simeq
m_{D,i}^2\left(1-\sum_{s=1}^{N}\frac{m_{D,i}^2}{M_s^2}\right),
\label{eq:lightpert}
\end{equation}
while the active component of shell $s$ is $m_{D,i}/M_s$ at leading order.
Thus the heavy-shell mixing is
\begin{equation}
\theta_{is}^2\simeq
\frac{m_{D,i}^2}{M_s^2}
=\frac{m_{D,i}^2}{8\mu^2s}.
\label{eq:thetashell}
\end{equation}

\begin{figure}[t]
\centering
\includegraphics[width=1.1\textwidth]{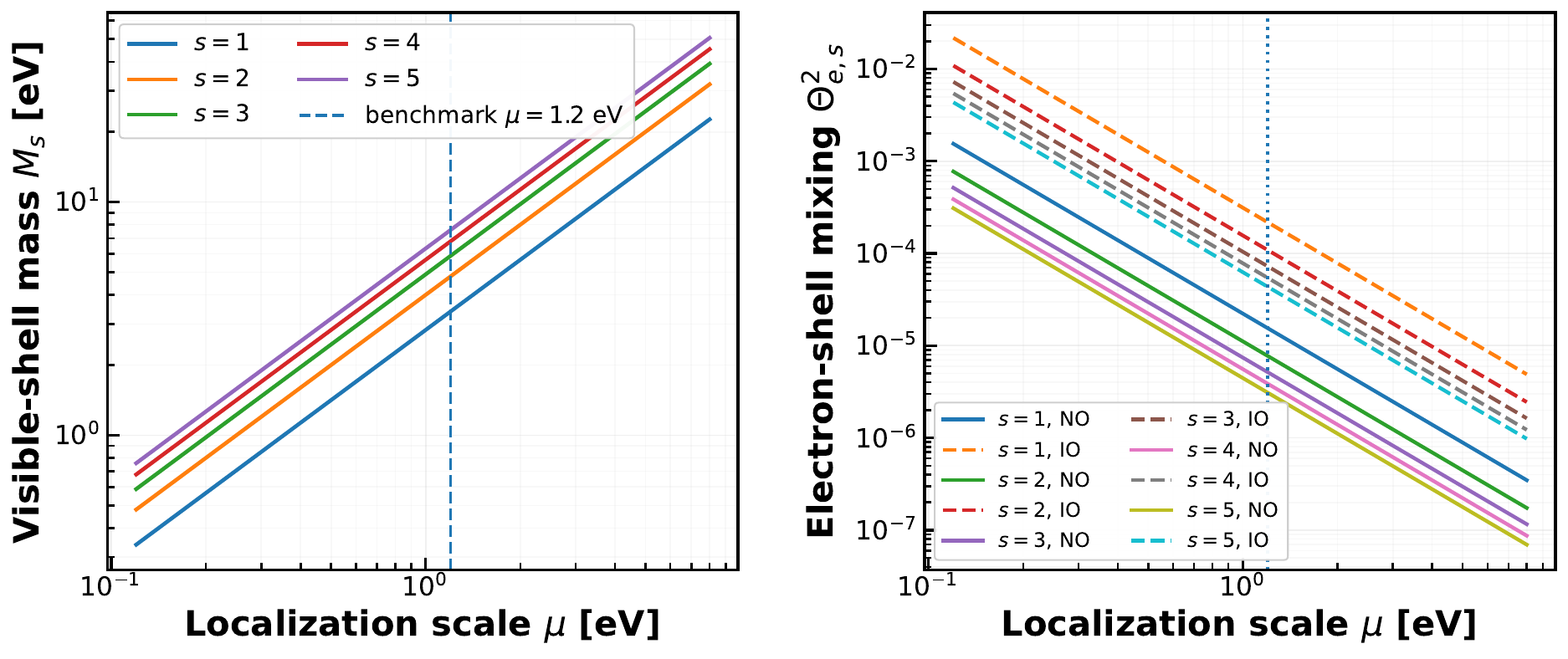}
\caption{\textcolor{black}{Localized sterile spectrum and brane mixing.
Left: masses of the first five brane-visible shells as functions of the
localization scale \(\mu\). Right: total electron-flavour mixing
\(\Theta_{e,s}^2\) for the same shells and
\(m_{\rm lightest}=0.01~\mathrm{eV}\); solid curves correspond to normal
ordering and dashed curves to inverted ordering.  The characteristic
relations \(M_s=\sqrt{8s}\,\mu\) and
\(\Theta_{e,s}^2\propto1/(\mu^2s)\) are shown directly.}}
\label{fig:shells}
\end{figure}

Summing over shells, the total active--sterile mixing from the localized
tower is
\begin{equation}
\eta_i(N)\equiv\sum_{s=1}^{N}\theta_{is}^2
=\frac{m_{D,i}^2}{8\mu^2}\sum_{s=1}^{N}\frac{1}{s}
=\frac{m_{D,i}^2}{8\mu^2}H_N,
\label{eq:etaexact}
\end{equation}
where $H_N=\sum_{s=1}^{N}1/s$ is the harmonic number.  For $N\gg1$,
\begin{equation}
H_N=\ln N+\gamma+\mathcal O(1/N),
\end{equation}
with $\gamma\simeq0.5772$ the Euler--Mascheroni constant.  The sum therefore grows only logarithmically with the number of shells,
rather than linearly or as a power.
The logarithmic growth means that ultraviolet information enters only
logarithmically.  Since the microscopic localizer and brane profile have not
been specified, we keep this ultraviolet dependence model independent and
parameterize it by the validity scale \(\Lambda_{\rm loc}\) introduced
below.

The harmonic description cannot be extrapolated to arbitrarily high excitation number. We therefore introduce a physical validity scale $\Lambda_{\rm loc}$ at which the local trapping potential departs from Eq.~\eqref{eq:Hperp}. Only shells with $M_s\lesssim\Lambda_{\rm loc}$ should be retained, giving
\begin{equation}
N_{\max}\simeq\frac{\Lambda_{\rm loc}^2}{8\mu^2}.
\end{equation}
The accumulated admixture is then
\begin{equation}
\eta_i(\Lambda_{\rm loc})\simeq
\frac{m_{D,i}^2}{8\mu^2}
\left[
\ln\left(\frac{\Lambda_{\rm loc}^2}{8\mu^2}\right)+\gamma_E
\right].
\label{eq:etalog}
\end{equation}
This logarithm is the localized counterpart of the classical codimension-two running discussed in Ref.~\cite{Dudas2005}. Because the ultraviolet localizer has not been specified, $\Lambda_{\rm loc}$ is not predicted by the effective theory. 
For numerical illustration in the following we take $\Lambda_{\rm loc}=M_6$ as an
illustrative maximal-cutoff choice within the effective description.  This
identification is not predicted by the harmonic localizer and should not be
interpreted as a model-independent ultraviolet bound.
A lower turnover scale reduces $\eta_i$, whereas the low-shell masses and individual mixings are unchanged.
A useful comparison is provided by five-dimensional Majorana towers, where
the presence of a bulk Majorana mass can suppress the active--sterile
admixture more strongly than in the localized Dirac case.  In the specific
construction of Ref.~\cite{AnchordoquiVafa2026}, the mixing of a mode with
mass \(m_n\) scales as
\(\theta_n\simeq m_D/m_n\), and the summed active--sterile admixture is
ultraviolet finite,
\[
\sum_n\theta_n^2\simeq\frac{m_\nu}{m_\psi}.
\]
By contrast, in the localized 2DD Dirac construction the brane-visible
spectrum satisfies \(M_s\propto\sqrt{s}\), so the corresponding unweighted
shell sum retains only logarithmic sensitivity to the localizer validity
scale, as in Eq.~\eqref{eq:etalog}. 

For thermal production, one must additionally weight the populated shells by their energy and include the finite-temperature dynamics of the
 active–sterile system.

The light singular value and its active normalization are, to leading order,
\begin{align}
 m_i^{\rm phys}&\simeq m_{D,i}\left(1-\frac{\eta_i}{2}\right),
 \label{eq:lightmasscorr}\\
 |V_{i0}|^2&\simeq1-\eta_i,
 \label{eq:lightnorm}
\end{align}
while $|V_{is}|^2\simeq\theta_{is}^2$.  Whenever $\eta_i\ll1$, Eq.~\eqref{eq:lightmasscorr} is only a small renormalization of the bare Yukawa parameter. We do not use the perturbative expansion outside that regime.

\section{Three-flavour embedding}
\label{sec:flavour} 
We assume that the localization background is flavour universal in the
sterile-generation space (or, equivalently, diagonal in the same sterile
basis).  The sterile fields may then be rotated by the right-handed unitary
matrix of the Yukawa singular-value decomposition without inducing
off-diagonal localization or tower mixing.
We work in the basis where the higher-dimensional Yukawa matrix is written
in singular-value form,
\begin{equation}
Y_6=U_{\rm PMNS}\,\mathrm{diag}(y_1,y_2,y_3),
\label{eq:Ydiag}
\end{equation}
so that each singular value $y_i$ couples one active mass combination  with an
independent localized sterile tower.  This is the same factorization used in
conventional large-extra-dimension (LED) studies~\cite{Forero2022}, where
the Standard Model is localized on a brane and sterile states propagate in
the bulk.  The difference here is that the bulk sterile states are localized near the brane rather than having a flat profile, 
so each tower consists of a massless zero mode plus the visible shell states derived in section  ~\ref{sec:spectrum}, with shell masses $M_s=\sqrt{8s}\,\mu$ and a shell-independent brane coupling $m_{D,i}$.
In this basis the flavour state is
\begin{equation}
|\nu_\alpha\rangle
=\sum_{i=1}^3U_{\alpha i}
\left[
V_{i0}|\nu_{i0}\rangle+\sum_{s\ge1}V_{is}|\nu_{is}\rangle
\right],
\label{eq:flavourstate}
\end{equation}
where $V_{i0}$ is the zero-mode component of the light mass eigenstate $i$
and $V_{is}$ is its component in the shell $s$.  The coefficients are
determined by diagonalizing the star matrix~\eqref{eq:starmatrix} for each
$i$. This factorized flavour structure is the direct analogue of the usual LED treatment: it separates the flavour mixing, encoded in $U_{\rm PMNS}$, from the geometric mixing into the tower, encoded in $V_{i0}$ and $V_{is}$.
It is therefore sufficient to isolate the geometric effect of localization.
For normal ordering,
\begin{align}
 m_1&=m_{\rm lightest},\\
 m_2&=\sqrt{m_1^2+\Delta m_{21}^2},\\
 m_3&=\sqrt{m_1^2+\Delta m_{31}^2},
\end{align}
while for inverted ordering we take
\begin{align}
 m_3&=m_{\rm lightest},\\
 m_2&=\sqrt{m_3^2+|\Delta m_{32}^2|},\\
 m_1&=\sqrt{m_2^2-\Delta m_{21}^2}.
\end{align}
We use the NuFIT~6.0 oscillation parameters \cite{NuFit2024}.  The numerical illustrations take
\begin{equation}
\sin^2\theta_{12}=0.307,\qquad
\sin^2\theta_{13}=0.02195,
\end{equation}
for the representative NO input, together with
\begin{equation}
\Delta m_{21}^2=7.49\times10^{-5}~\mathrm{eV}^2,
\qquad
\Delta m_{31}^2=2.534\times10^{-3}~\mathrm{eV}^2,
\end{equation}
and $|\Delta m_{32}^2|=2.510\times10^{-3}~\mathrm{eV}^2$ for IO.  The small ordering-dependent shifts of the mixing angles are included in the numerical code.  Unless otherwise stated, the plots take $m_{\rm lightest}=0.01~\mathrm{eV}$.
Neglecting the small correction in Eq.~\eqref{eq:lightmasscorr}, the Yukawa eigenvalues required for the measured masses are
\begin{equation}
y_i\simeq
\sqrt\pi\frac{M_6}{v}
\frac{m_i}{\mu}.
\label{eq:yifit}
\end{equation}
This relation makes the origin of the improvement transparent: in the
localized theory the light Dirac mass is controlled by the brane value of
the trapped zero mode, \(m_D^{\rm loc}\propto\mu/M_6\), rather than by the
flat-volume factor \(V_2^{-1/2}\).
For the benchmark
\begin{equation}
M_6=10~\mathrm{TeV},
\qquad
\mu=1.2~\mathrm{eV},
\qquad
m_{\rm lightest}=0.01~\mathrm{eV},
\label{eq:benchmark}
\end{equation}
we obtain
\begin{equation}
(y_1,y_2,y_3)_{\rm NO}
\simeq(0.60,0.79,3.08),
\label{eq:yNO}
\end{equation}
and
\begin{equation}
(y_1,y_2,y_3)_{\rm IO}
\simeq(3.02,3.07,0.60).
\label{eq:yIO}
\end{equation}
All of these values are perturbative.  At the same point
\begin{equation}
\mu R\simeq4.65,
\qquad
\xi_{\rm loc}\simeq23.3,
\end{equation}
so the localized wave function supplies the enhancement that would otherwise have to come from a large Yukawa coupling.

\section{Oscillation phenomenology}
\label{sec:osc}

In vacuum the flavour-transition amplitude is
\begin{equation}
\mathcal A_{\alpha\beta}(L,E)
=\sum_{i=1}^3U_{\alpha i}U_{\beta i}^*
\left[
|V_{i0}|^2e^{-i m_i^2L/(2E)}
+\sum_{s\ge1}|V_{is}|^2
 e^{-i M_{is}^2L/(2E)}
\right],
\label{eq:oscamp}
\end{equation}
with
\begin{equation}
P_{\alpha\beta}=|\mathcal A_{\alpha\beta}|^2.
\end{equation}
Here $M_{is}^2$ differs from $M_s^2$ only by corrections of order $m_i^2$ that are negligible for the heavy shells.
For reactor antineutrinos the observable of primary interest is
\begin{equation}
P_{ee}=|\mathcal A_{ee}|^2.
\end{equation}
The first visible localized splitting is
\begin{equation}
\Delta m_{\rm loc,1}^2\simeq8\mu^2.
\label{eq:dmloc}
\end{equation}
For $\mu=1.2~\mathrm{eV}$ this is approximately $11.5~\mathrm{eV}^2$, so the new phase oscillates many times even at the Daya Bay near-hall baselines \cite{DayaBay2023}.  The localized shells are therefore effectively averaged in reactor spectra.
After averaging the heavy phases, the leading coherent part becomes
\begin{equation}
P_{ee}^{\rm av}
\simeq
\left|
\sum_i |U_{ei}|^2(1-\eta_i)
 e^{-i m_i^2L/(2E)}
\right|^2
+\mathcal O(\theta^4).
\label{eq:Peeavg}
\end{equation}
The leading effect is thus a small electron-row normalization shift plus a tiny change in the relative standard-oscillation amplitudes.  This is qualitatively different from a flat 2DD tower with first mass near $R^{-1}$, where the low KK modes remain dynamically resolved at reactor baselines.

\begin{figure}[t]
\centering
\includegraphics[width=0.82\textwidth]{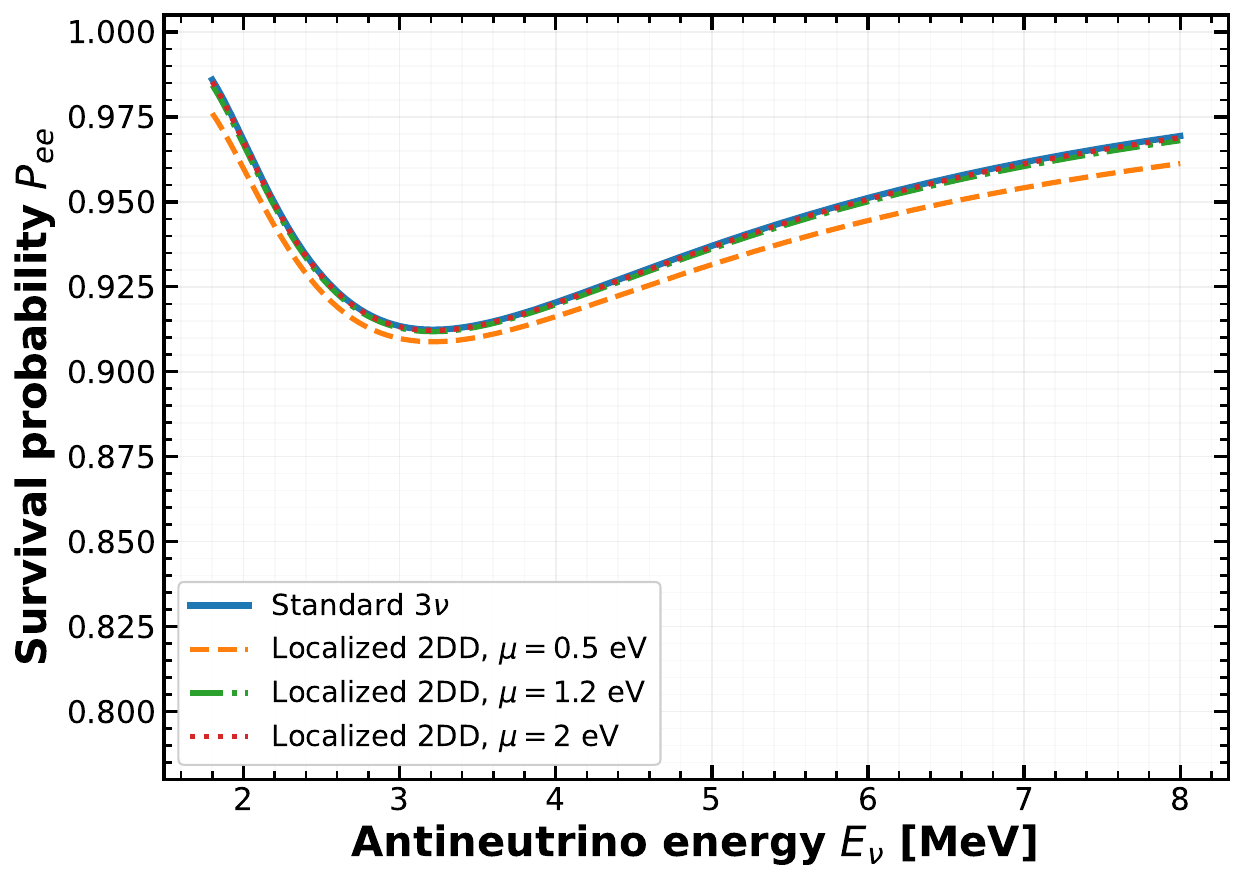}
\caption{Electron-antineutrino survival probability for the standard three-neutrino case and representative localized-2DD benchmarks at $L=1.6~\mathrm{km}$, for normal ordering and $m_{\rm lightest}=0.01~\mathrm{eV}$.  The heavy localized shells are averaged according to Eq.~\eqref{eq:Peeavg}.  This figure is the localized-2DD counterpart of the survival-probability comparison used in recent dark-dimension analyses; it is illustrative rather than a replacement for a full Daya Bay near/far likelihood.}
\label{fig:peediff}
\end{figure}

Figure~\ref{fig:peediff} shows the electron-antineutrino survival probability itself, rather than only a residual, following the observable used in recent dark-dimension oscillation analyses.  The curves make the physical point transparent: once the first localized shell is sufficiently heavy, the rapid sterile phases are averaged and the surviving energy dependence stays close to the standard three-neutrino prediction.  For $\mu\gtrsim1~\mathrm{eV}$ the distortion is small while the gravitational radius can remain near the value dictated by the 2DD relation.
Define
\begin{equation}
{\cal N}_e
\equiv\sum_{i=1}^3 |U_{ei}|^2|V_{i0}|^2
\simeq1-\eta_e,
\end{equation}
where
\begin{equation}
\eta_e=\sum_i|U_{ei}|^2\eta_i.
\label{eq:etae}
\end{equation}
Using Eq.~\eqref{eq:etalog},
\begin{equation}
\eta_e\simeq
\frac{m_\beta^2}{8\mu^2}
\left[
\textcolor{black}{
\ln\left(\frac{\Lambda_{\rm loc}^2}{8\mu^2}\right)}
+\gamma_E
\right],
\label{eq:etaembeta}
\end{equation}

where
\begin{equation}
m_\beta^2\equiv\sum_i|U_{ei}|^2m_i^2.
\label{eq:mbeta}
\end{equation}
Thus, the electron-row normalization is directly related to the same effective mass that appears in beta decay.

A recent combined non-unitarity analysis using reactor and atmospheric data finds electron-row normalization constrained at the few-percent level \cite{Kozynets2024}.  We use this only as an order-of-magnitude cross-check because a tower of eV-scale states is not identical to a generic non-unitary mixing matrix.  For the benchmark $m_{\rm lightest}=0.01~\mathrm{eV}$, $\mu=1.2~\mathrm{eV}$ and the  illustrative maximal-cutoff choice $\Lambda_{\rm loc}=M_6=10~\mathrm{TeV}$, we find
\begin{equation}
\eta_e\simeq8.97\times10^{-4}\quad(\mathrm{NO}),
\qquad
\eta_e\simeq1.26\times10^{-2}\quad(\mathrm{IO}).
\label{eq:etaebench}
\end{equation}
These values indicate that the benchmark is naturally much closer to the three-neutrino limit in NO than in IO.  They are not experimental confidence limits; a formal reactor exclusion requires the localized tower to be propagated through the measured baselines, detector response and covariance information.
It is useful to state explicitly why Eq.~\eqref{eq:flatbounds} cannot simply be imposed on the localized theory.  The flat LED analysis assumes that the sterile masses are set by
\begin{equation}
M_{\bm n}^{\rm flat}=\frac{|\bm n|}{R},
\end{equation}
and that the same brane operator generates the zero-mode mass and the couplings to the KK tower \cite{Elacmaz2025}.  In our localized theory, the measured light mass is instead controlled by the brane value of the trapped zero mode, Eq.~\eqref{eq:mDloc}, while the visible sterile spectrum begins at
\begin{equation}
M_{\rm vis,1}=2\sqrt2\,\mu.
\end{equation}
For $\mu\gg R^{-1}$ these are parametrically different scales.  The reactor experiment therefore probes $\mu$ and the active normalization rather than directly measuring the gravitational compactification radius.  This decoupling is the principal reason the $R\sim0.7$--$1~\mu\mathrm m$ gravitational region can coexist with a neutrino sector that would be excluded if the sterile wave function were flat.

\section{Beta decay and localized-shell kinks}
\label{sec:beta}

Tritium beta decay is a complementary probe because it is sensitive to neutrino masses kinematically and does not rely on the oscillation phase.  KATRIN currently constrains the effective electron-neutrino mass to $m_\beta<0.45~\mathrm{eV}$ at 90\% C.L. \cite{KATRINmass2025}.  The same data have also been used for a direct sterile-neutrino search over the last $40~\mathrm{eV}$ below the endpoint \cite{KATRINsterile2025}; the relevant spectral and response treatment is described in Ref.~\cite{Kleesiek2019}.  Extra-dimensional sterile states have long been recognized as producing kinematic distortions in beta decay \cite{McLaughlinNg2001,Basto2013,AntoniadisKATRIN2026}, while the TRISTAN program extends the reach toward heavier sterile states \cite{TRISTAN2019,TRISTAN2026}.
For a set of mass eigenstates the neutrino-dependent part of the differential beta spectrum is
\begin{equation}
S(q)=\sum_a |U_{ea}|^2
q\sqrt{q^2-m_a^2}\,\Theta(q-m_a),
\qquad q\equiv E_0-K_e.
\label{eq:betaspec}
\end{equation}
In our localized theory, the total electron mixing carried by shell $s$ is, at leading order,
\begin{equation}
\Theta_{e,s}^2
\equiv\sum_i|U_{ei}|^2\theta_{is}^2
=\frac{m_\beta^2}{8\mu^2s}.
\label{eq:betashellmix}
\end{equation}
The kink positions are
\begin{equation}
M_s=\sqrt{8s}\,\mu,
\qquad
\Theta_{e,s}^2=\frac{m_\beta^2}{8\mu^2s}.
\label{eq:betaprediction}
\end{equation}
This pair of equations is a characteristic prediction of the Gaussian localizer benchmark: the thresholds grow as $\sqrt{s}$, while the squared electron-flavour mixing (equivalently, the shell admixture) decreases as $1/s$.
For the benchmark $\mu=1.2~\mathrm{eV}$, the first visible kink lies at
\begin{equation}
M_1\simeq3.39~\mathrm{eV}.
\end{equation}
For $m_{\rm lightest}=0.01~\mathrm{eV}$ the first-shell electron mixing is
\begin{equation}
\Theta_{e,1}^2\simeq1.55\times10^{-5}\quad(\mathrm{NO}),
\qquad
\Theta_{e,1}^2\simeq2.17\times10^{-4}\quad(\mathrm{IO}).
\label{eq:firstshellnumbers}
\end{equation}
The latest KATRIN sterile-neutrino analysis searches for spectral kinks over the final $40~\mathrm{eV}$ below the tritium endpoint and constrains sterile admixtures over a broad eV-scale mass range \cite{KATRINsterile2025}.  Our benchmark therefore lies in the kinematic region to which KATRIN is sensitive.  A one-shell comparison is useful for orientation, but the localized model predicts several thresholds simultaneously.  A statistically meaningful limit must sum all shells inside the analysis window and fold the result with the molecular final-state distribution, transmission function, energy-loss model, detector response and nuisance parameters.

\begin{figure}[t]
\centering
\includegraphics[width=0.82\textwidth]{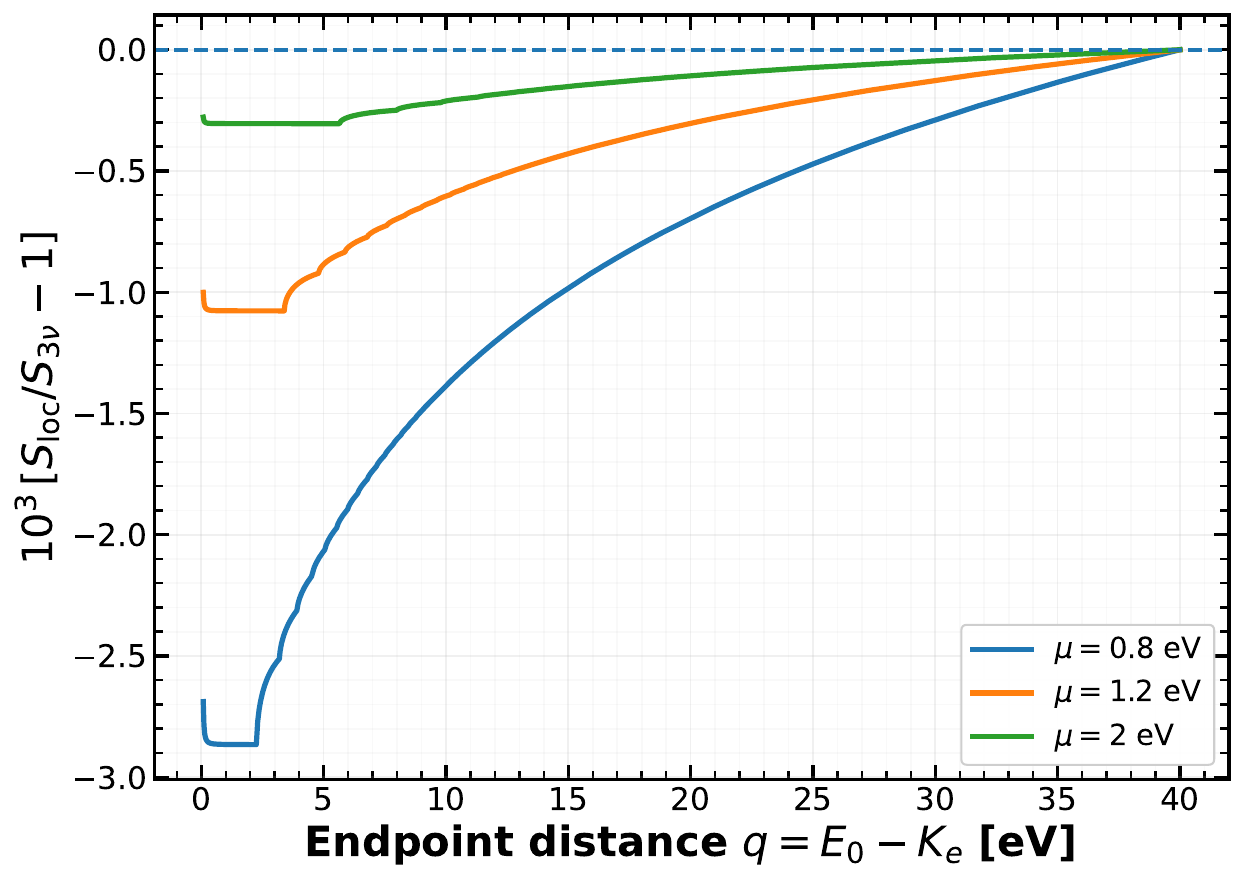}
\caption{Beta-endpoint shape distortion for inverted ordering in the localized 2DD benchmark.  The spectrum is normalized at $q=40~\mathrm{eV}$ because the overall signal normalization is floated in the experimental analysis.  Successive changes in curvature occur at $M_s=\sqrt{8s}\,\mu$.  The curves show the characteristic multi-threshold structure before convolution with the full KATRIN response.}
\label{fig:beta}
\end{figure}

The normalized spectral distortion for IO is illustrated in Fig.~\ref{fig:beta}.  The endpoint normalization is floated in the plot, mimicking the fact that KATRIN analyses fit an overall signal normalization.  The small staircase structure is generated by successive localized shells.
The IO spectrum is substantially more favorable for beta-decay searches because $m_\beta$ is larger.

\section{Relation to Majorana neutrinos and lepton-number violation}
\label{sec:majorana}

The construction developed here is a Dirac-neutrino model.  This choice is deliberate.  In six dimensions the Lorentz structure of fermion mass terms differs from the five-dimensional case, so the Majorana bilinear used in a 5D dark-dimension model cannot be imported unchanged.  A consistent 6D lepton-number-violating completion generally requires additional spinor structure, boundary conditions or a scalar sector that pairs the appropriate six-dimensional chiralities.  The phenomenology of neutrinoless double-beta decay in the presence of sterile or heavy neutral fermions depends sensitively on their masses, mixings, and Majorana structure \cite{Bhattacharyya2003,BlennowSeesaw2010,Rodejohann2011,Dolinski2019,Abada2019,Faessler2014,Bolton2025}.  The recent 5D analysis of Ref.~\cite{DeGiorgiPasari2026} illustrates why this distinction matters: the detailed Majorana structure reorganizes the KK spectrum and produces nontrivial screening in neutrinoless double beta decay.

For that reason we do not apply $0\nu\beta\beta$ limits to the present Dirac benchmark, despite the strong current experimental sensitivity to Majorana-neutrino exchange \cite{KamLANDZen2025}.  The localization mechanism, the resolution of the Yukawa problem and the oscillation/beta-decay signatures derived above do not require lepton-number violation.  A Majorana extension can be built by coupling a consistent 6D paired-spinor sector to the same localizer, at which point the $0\nu\beta\beta$ amplitude must be derived from the completed mass matrix rather than assumed by analogy with a 5D model.

\section{Numerical benchmarks}
\label{sec:numerics}

Table~\ref{tab:bench} summarizes the central benchmark used throughout the paper. The quoted $\eta_e$ values use Eq.~\eqref{eq:etaembeta} with the illustrative choice $\Lambda_{\rm loc}=M_6$; they should therefore be read as benchmark values rather than cutoff-independent predictions of the low-energy theory.
\begin{table}[t]
\centering
\caption{Localized 2DD benchmark with $M_6=10~\mathrm{TeV}$, $R=0.765~\mu\mathrm m$, $\mu=1.2~\mathrm{eV}$ and $m_{\rm lightest}=0.01~\mathrm{eV}$.}
\label{tab:bench}
\begin{tabular}{lcc}
\toprule
Quantity & NO & IO\\
\midrule
$(m_1,m_2,m_3)$ [eV] & $(0.0100,0.0132,0.0513)$ & $(0.05035,0.05109,0.0100)$\\
$(y_1,y_2,y_3)$ & $(0.600,0.794,3.082)$ & $(3.023,3.067,0.600)$\\
$m_\beta$ [eV] & $0.01335$ & $0.05004$\\
$\eta_e$ & $8.97\times10^{-4}$ & $1.26\times10^{-2}$\\
$M_{\rm vis,1}$ [eV] & $3.394$ & $3.394$\\
$\Theta_{e,1}^2$ & $1.55\times10^{-5}$ & $2.17\times10^{-4}$\\
\bottomrule
\end{tabular}
\end{table}
Several observations follow readily.

First, the localization scale required to restore perturbative Yukawa couplings is modest compared with the fundamental scale.  It is an infrared parameter of the transverse fermion potential, not a new gravitational scale.  For the benchmark,
\begin{equation}
\mu R\simeq4.65,
\end{equation}
so the sterile zero mode is localized within a region much smaller than the compactification radius while still being enormously larger than the fundamental length $M_6^{-1}$.

Second, NO is comparatively easy to accommodate because the electron flavour is dominated by the two lighter mass eigenstates.  In IO, the electron flavour overlaps strongly with two states near $0.05~\mathrm{eV}$, so $m_\beta$ and $\eta_e$ are larger.  This explains why both flat LED reactor bounds and the localized normalization test are more restrictive in IO.

Third, the relevant sterile mass gap is no longer $R^{-1}$.  For the $10~\mathrm{TeV}$ gravity benchmark,
\begin{equation}
R^{-1}\simeq0.258~\mathrm{eV},
\end{equation}
while
\begin{equation}
M_{\rm vis,1}\simeq3.39~\mathrm{eV}.
\end{equation}
This order-of-magnitude separation is enough to move the additional oscillations from the partially resolved reactor regime into the rapid-averaging regime.

The main physical point of the localized construction is that the gravitational radius and the neutrino excitation scale need not be identical.  In the flat theory, the same radius $R$ fixes both the wave-function normalization and the KK gap, so increasing the light neutrino mass through a larger brane Yukawa coupling also strengthens the coupling to a dense two-dimensional tower.  Localization separates these roles.  The compactification radius continues to satisfy the 2DD gravity relation, whereas the brane value of the sterile zero mode and the first visible sterile mass are controlled by the dimensionless localization parameter $\kappa_{\rm loc}=\mu R$.

The comparison with conventional LED analyses is therefore informative but must be made at the level of physical masses and overlaps rather than by transplanting radius bounds.  In the localized benchmark the first visible state appears at $M_1=2\sqrt2\,\mu$, and the squared shell mixing falls as $1/s$.  The mass- and mixing-spectrum presentation in Fig.~\ref{fig:shells} is the direct analogue of the quantities used in standard LED fits such as Ref.~\cite{Forero2022}, but the mapping from $R$ to the laboratory spectrum is different.  This is why the published flat-LED limits in Fig.~\ref{fig:rbounds} serve as experimental reference scales rather than direct exclusions of the present model.

The effective description also makes clear where ultraviolet information enters.  The low-lying results depend only on the localized profile and are reliable for $\mu R\gg1$, whereas the logarithmic sum over highly excited shells depends on the turnover scale $\Lambda_{\rm loc}$ of the microscopic trapping potential.
We therefore use $\Lambda_{\rm loc}=M_6$ only as an illustrative maximal-cutoff
choice within the effective description; the actual turnover scale is a
property of the microscopic localization mechanism.
A finite brane thickness or an explicit scalar/vortex completion would regulate the high shells without changing the first few masses and overlaps that control laboratory phenomenology.

Cosmology provides an important but model-dependent test.  Sterile and bulk-neutrino populations can affect $N_{\rm eff}$ and the neutrino mass density \cite{Vincent2015,Anchordoqui2025,McKeen2024}.  If the eV-scale localized shells thermalize in a standard high-reheat cosmology, strong constraints are expected.  Their production rate, however, depends on the brane-bulk interactions, localization profile, reheating temperature and ultraviolet completion.  We therefore do not apply flat-tower cosmological limits directly; a consistent cosmological assessment requires the corresponding production and Boltzmann equations for the localized spectrum.

More generally, the cosmological production of bulk sterile-neutrino towers
suggests a useful scaling diagnostic that can be applied directly to our
localized shell spectrum before performing a full Boltzmann analysis
\cite{Abazajian2003,AnchordoquiVafa2026}.  Collision-induced production of
a sterile mode is parametrically controlled by
\(\Gamma_{\rm act}\langle P_{\rm active\to sterile}\rangle\), and in the
regime where finite-temperature matter effects are subdominant this scales
as \(G_F^2T^5\theta^2\).  The number of brane-visible localized shells below
an energy \(E\) is
\begin{equation}
N_{\rm vis}(E)
\simeq
\frac{E^2}{8\mu^2},
\qquad
E\equiv\min(T_{\rm max},\Lambda_{\rm loc}),
\label{eq:Nthermal}
\end{equation}
where \(T_{\rm max}\) denotes the largest temperature at which the
four-dimensional effective description and the active-plasma treatment are
being applied.  For one light-neutrino eigenstate, the energy-weighted
vacuum-mixing sum over the kinematically accessible shells is therefore
\begin{align}
{\cal W}_i(E)
&\equiv
\sum_{M_s<E}M_s\,\theta_{is}^2
=
\frac{m_i^2}{\sqrt{8}\,\mu}
\sum_{s=1}^{N_{\rm vis}(E)}\frac{1}{\sqrt{s}}
\nonumber\\
&\simeq
\frac{m_i^2}{4\mu^2}\,E ,
\qquad
N_{\rm vis}\gg1 .
\label{eq:thermalweight}
\end{align}
The last step uses
\(\sum_{s=1}^{N}s^{-1/2}\simeq2\sqrt N\).
Thus the laboratory quantity
\(\sum_s\theta_{is}^2\) grows only logarithmically, whereas the
energy-weighted vacuum-mixing proxy entering a simplified production estimate
grows linearly with the highest accessible shell mass in the harmonic regime.
This scaling is not, by itself, a prediction for the final cosmological
energy density.  Equivalently, the visible-shell density is
\(dN_{\rm vis}/dM=M/(4\mu^2)\), so the increasing two-dimensional density of
states compensates one power of the mixing suppression.

This observation does not by itself constitute a cosmological exclusion.
The in-medium mixing angles, the thermal history, the actual turnover of the
localizer, possible finite-width suppression of high-shell overlaps, and
additional interactions can all modify the production rate.  In the
five-dimensional Majorana realization of
Ref.~\cite{AnchordoquiVafa2026}, for example, the suppression mechanism is
controlled by a bulk Majorana mass, and the quoted
\(m_\psi\sim100~\mathrm{GeV}\) scale is specific to that setup and its assumed
thermal history.  It therefore cannot be directly transferred to the
localized Dirac spectrum considered here.  Nevertheless,
Eqs.~\eqref{eq:Nthermal}--\eqref{eq:thermalweight} show that cosmological
viability can be more sensitive to \(T_{\rm max}\) and
\(\Lambda_{\rm loc}\) than the zero-temperature non-unitarity measure alone
would suggest.  A dedicated finite-temperature Boltzmann analysis of the
localized shells is therefore an important next step.

The laboratory analysis is correspondingly conservative.  The oscillation curves display the characteristic averaged behavior and the beta-decay calculation displays the predicted multi-threshold structure, while published flat-LED and sterile-neutrino results are used as external reference scales.  Dedicated Daya Bay/RENO or KATRIN likelihoods would sharpen the numerical exclusion contours but would not alter the analytic relations $M_s^2=8\mu^2s$ and $\Theta_{e,s}^2=m_\beta^2/(8\mu^2s)$ that define the localized 2DD signal.

\section{Conclusions}

We have investigated a localized-neutrino realization within the
two-dark-dimensions framework. The motivation comes from the flat-bulk
construction, where the volume suppression of the sterile-neutrino zero
mode makes the effective Dirac mass too small for
$M_6\sim10~\mathrm{TeV}$ unless the higher-dimensional Yukawa coupling is
large, while the accompanying two-dimensional KK tower is strongly
constrained by oscillation data. Localizing the sterile wave function near
the Standard-Model brane provides a simple way to address both issues.

The main results of our analysis can be summarized as follows:
\begin{itemize}

\item
The localization mechanism enhances the sterile zero-mode wave function at
the SM brane and modifies the effective Dirac mass according to
Eq.~\eqref{eq:mDloc}. As a result, realistic neutrino masses can be obtained
with perturbative, $\mathcal O(1)$ Yukawa couplings instead of the large
coupling required in the flat realization, Eq.~\eqref{eq:mDflatnum}.

\item
The localized excitations are reorganized into brane-visible shells with
masses given by Eq.~\eqref{eq:shellmass}, while their active--sterile mixing
follows Eq.~\eqref{eq:thetashell}. The total sterile admixture grows only
logarithmically with the number of shells below the validity scale of the
localizer, as described by Eq.~\eqref{eq:etalog}.

\item
For the benchmark point of Eq.~\eqref{eq:benchmark}, with
$M_6=10~\mathrm{TeV}$, $R\simeq0.765~\mu\mathrm{m}$ and
$\mu=1.2~\mathrm{eV}$, the required Yukawa eigenvalues are of
$\mathcal O(1)$. The corresponding electron-flavour sterile admixture is
below $10^{-3}$ for normal ordering and at the percent level for inverted
ordering, Eq.~\eqref{eq:etaebench}. The first brane-visible state lies near
$3.4~\mathrm{eV}$, so the additional oscillations are rapidly averaged at
reactor baselines.

\item
The localized construction predicts a characteristic multi-threshold
structure in beta decay, with shell masses and electron-flavour mixings
related by Eq.~\eqref{eq:betaprediction}. This provides a complementary
probe of the model alongside oscillation measurements. Since the sterile
spectrum is controlled by the localization scale rather than directly by
$R^{-1}$, bounds obtained for flat LED models cannot be directly applied to
the localized theory.

\end{itemize}

The central consequence of localization is therefore the separation of the
gravitational compactification scale from the experimentally relevant
sterile-neutrino scale, while preserving a predictive relation between the
light-neutrino masses, localized-shell masses and active--sterile mixing.
A complete experimental test will require dedicated reactor and KATRIN
likelihood analyses including the full localized tower.

An additional consequence is that laboratory and cosmological observables
weight this tower differently. While the zero-temperature sterile admixture
grows only logarithmically with the number of visible shells, the
corresponding energy-weighted vacuum-mixing sum grows linearly with the
highest accessible shell mass in the harmonic regime,
Eq.~\eqref{eq:thermalweight}. This is a scaling diagnostic rather
than a prediction of the final thermal abundance. Cosmological viability therefore depends
sensitively on the thermal history and on the ultraviolet turnover scale of
the localizer, and requires a dedicated finite-temperature treatment.
Likewise, neutrinoless double-beta decay requires a consistent
six-dimensional source of lepton-number violation. These issues provide
natural directions for extending the present framework.

\section*{Code availability}

The numerical codes and Jupyter notebooks used to reproduce the figures and
benchmark results presented in this work are publicly available at
\url{https://github.com/WaqasAhmedmit/localized-neutrinos-2dd}.


\begin{thebibliography}{99}

\bibitem{Dienes1999}
K.~R.~Dienes, E.~Dudas and T.~Gherghetta,
``Neutrino oscillations without neutrino masses or heavy mass scales: A higher dimensional seesaw mechanism,''
Nucl. Phys. B \textbf{557} (1999) 25,
\href{https://arxiv.org/abs/hep-ph/9811428}{arXiv:hep-ph/9811428}.

\bibitem{ArkaniHamed2001}
N.~Arkani-Hamed, S.~Dimopoulos, G.~R.~Dvali and J.~March-Russell,
``Neutrino masses from large extra dimensions,''
Phys. Rev. D \textbf{65} (2001) 024032,
\href{https://arxiv.org/abs/hep-ph/9811448}{arXiv:hep-ph/9811448}.

\bibitem{DvaliSmirnov1999}
G.~R.~Dvali and A.~Y.~Smirnov,
``Probing large extra dimensions with neutrinos,''
Nucl. Phys. B \textbf{563} (1999) 63--81,
\href{https://arxiv.org/abs/hep-ph/9904211}{arXiv:hep-ph/9904211}.

\bibitem{Lukas2001}
A.~Lukas, P.~Ramond, A.~Romanino and G.~G.~Ross,
``Neutrino Masses and Mixing in Brane World Theories,''
JHEP \textbf{04} (2001) 010,
\href{https://arxiv.org/abs/hep-ph/0011295}{arXiv:hep-ph/0011295}.

\bibitem{KimRaby2003}
H.~D.~Kim and S.~Raby,
``Neutrinos in 5-D SO(10) unification,''
JHEP \textbf{07} (2003) 014,
\href{https://arxiv.org/abs/hep-ph/0304104}{arXiv:hep-ph/0304104}.

\bibitem{EiseleHaba2006}
M.-T.~Eisele and N.~Haba,
``Family symmetry and single right-handed neutrino dominance in five dimensions,''
Phys. Rev. D \textbf{74} (2006) 073007,
\href{https://arxiv.org/abs/hep-ph/0603158}{arXiv:hep-ph/0603158}.

\bibitem{WatanabeYoshioka2011}
A.~Watanabe and K.~Yoshioka,
``Seesaw in the bulk,''
Prog. Theor. Phys. \textbf{125} (2011) 129--148,
\href{https://arxiv.org/abs/1007.1527}{arXiv:1007.1527}.

\bibitem{DiegoQuiros2008}
D.~Diego and M.~Quiros,
``Dirac Versus Majorana Neutrino Masses From a TeV Interval,''
Nucl. Phys. B \textbf{805} (2008) 148--167,
\href{https://arxiv.org/abs/0804.2838}{arXiv:0804.2838}.

\bibitem{Blennow2010}
M.~Blennow, H.~Melbeus, T.~Ohlsson and H.~Zhang,
``Signatures from an extra-dimensional seesaw model,''
Phys. Rev. D \textbf{82} (2010) 045023,
\href{https://arxiv.org/abs/1003.0669}{arXiv:1003.0669}.

\bibitem{MonteroVafa2023}
M.~Montero, C.~Vafa and I.~Valenzuela,
``The dark dimension and the Swampland,''
JHEP \textbf{02} (2023) 022,
\href{https://arxiv.org/abs/2205.12293}{arXiv:2205.12293}.

\bibitem{Antoniadis1990}
I.~Antoniadis,
``A Possible new dimension at a few TeV,''
Phys. Lett. B \textbf{246} (1990) 377--384.

\bibitem{ADD1998}
N.~Arkani-Hamed, S.~Dimopoulos and G.~R.~Dvali,
``The Hierarchy problem and new dimensions at a millimeter,''
Phys. Lett. B \textbf{429} (1998) 263--272,
\href{https://arxiv.org/abs/hep-ph/9803315}{arXiv:hep-ph/9803315}.

\bibitem{AntoniadisADD1998}
I.~Antoniadis, N.~Arkani-Hamed, S.~Dimopoulos and G.~R.~Dvali,
``New dimensions at a millimeter to a Fermi and superstrings at a TeV,''
Phys. Lett. B \textbf{436} (1998) 257--263,
\href{https://arxiv.org/abs/hep-ph/9804398}{arXiv:hep-ph/9804398}.

\bibitem{Lee2020}
J.~G.~Lee, E.~G.~Adelberger, T.~S.~Cook, S.~M.~Fleischer and B.~R.~Heckel,
``New Test of the Gravitational $1/r^2$ Law at Separations down to 52 microns,''
Phys. Rev. Lett. \textbf{124} (2020) 101101,
\href{https://arxiv.org/abs/2002.11761}{arXiv:2002.11761}.

\bibitem{Murata2026}
J.~Murata, T.~Fujiie and S.~Suzuki,
``Short-Range Tests of the Gravitational Inverse-Square Law,''
\href{https://arxiv.org/abs/2605.18212}{arXiv:2605.18212} (2026).

\bibitem{Machado2011}
P.~A.~N.~Machado, H.~Nunokawa and R.~Zukanovich Funchal,
``Testing for Large Extra Dimensions with Neutrino Oscillations,''
Phys. Rev. D \textbf{84} (2011) 013003,
arXiv:1101.0003 [hep-ph].

\bibitem{Machado2012}
P.~A.~N.~Machado, H.~Nunokawa, F.~A.~P.~dos Santos and R.~Zukanovich Funchal,
Phys. Rev. D \textbf{85} (2012) 073012,
arXiv:1107.2400 [hep-ph].

\bibitem{Esmaili2014}
A.~Esmaili, O.~L.~G.~Peres and Z.~Tabrizi,
``Probing Large Extra Dimensions With IceCube,''
JCAP \textbf{12} (2014) 002,
arXiv:1409.3502 [hep-ph].

\bibitem{GirardiMeloni2014}
I.~Girardi and D.~Meloni,
Phys. Rev. D \textbf{90} (2014) 073011,
arXiv:1403.5507 [hep-ph].

\bibitem{Stenico2018}
G.~V.~Stenico, D.~V.~Forero and O.~L.~G.~Peres,
``A Short Travel for Neutrinos in Large Extra Dimensions,''
JHEP \textbf{11} (2018) 155,
arXiv:1808.05450 [hep-ph].

\bibitem{BastoJUNO2022}
V.~S.~Basto-Gonzalez, D.~V.~Forero, C.~Giunti, A.~A.~Quiroga and C.~A.~Ternes,
``Short-baseline oscillation scenarios at JUNO and TAO,''
Phys. Rev. D \textbf{105} (2022) 075023,
arXiv:2112.00379 [hep-ph].

\bibitem{Forero2022}
D.~V.~Forero, C.~Giunti, C.~A.~Ternes and O.~Tyagi,
``Large extra dimensions and neutrino experiments,''
Phys. Rev. D \textbf{106} (2022) 035027,
\href{https://arxiv.org/abs/2207.02790}{arXiv:2207.02790}.

\bibitem{Eller2025}
P.~Eller, M.~Ettengruber and A.~Zander,
``Neutrino data analysis of extra-dimensional theories with massive bulk fields,''
Phys. Rev. D \textbf{112} (2025) 055009,
\href{https://arxiv.org/abs/2508.04274}{arXiv:2508.04274}.

\bibitem{DeGiorgiTurner2026}
A.~de Giorgi, D.~Pasari and J.~Turner,
``Do neutrinos dream in 5D? Towards a comprehensive extra-dimensional neutrino phenomenology,''
JHEP \textbf{05} (2026) 152,
\href{https://arxiv.org/abs/2512.02101}{arXiv:2512.02101}.

\bibitem{AntoniadisKATRIN2026}
I.~Antoniadis, A.~Chatrabhuti and H.~Isono,
``Searching for a dark dimension right-handed neutrino in KATRIN,''
JHEP \textbf{02} (2026) 015,
\href{https://arxiv.org/abs/2509.05233}{arXiv:2509.05233}.

\bibitem{Bai2026}
A.-Y.~Bai, A.~Chatrabhuti, Y.-Y.~Huang, H.~Isono and J.~Tang,
``Dark Dimension Right-handed Neutrinos Confronted with Long-Baseline Oscillation Experiments,''
JHEP \textbf{05} (2026) 196,
\href{https://arxiv.org/abs/2601.00790}{arXiv:2601.00790}.

\bibitem{Franklin2026}
J.~Franklin, R.~Kamath, D.~Pasari, Y.~F.~Perez-Gonzalez, J.~Turner and M.~A.~Vogiatzi,
``nuSTORM as a Precision Probe of the Standard Model and New Physics,''
JHEP \textbf{02} (2026) 127,
\href{https://arxiv.org/abs/2509.08882}{arXiv:2509.08882}.

\bibitem{MonteroBminusL2026}
M.~Montero, C.~Vafa and I.~Valenzuela,
``Neutrinos, $B-L$ symmetry and the dark dimension,''
JHEP \textbf{06} (2026) 059,
\href{https://arxiv.org/abs/2512.09052}{arXiv:2512.09052}.

\bibitem{Abe2026}
Y.~Abe, Y.~Adachi and Y.~Fujimoto,
``Conjugate Boundary Conditions, Kaluza-Klein Fermions, and an Extended Seesaw Model,''
\href{https://arxiv.org/abs/2607.11150}{arXiv:2607.11150} (2026).

\bibitem{Lonardi2026}
S.~Lonardi, M.~Ettengruber and P.~Eller,
``Searching for the N-naturalness tower of neutrinos,''
\href{https://arxiv.org/abs/2607.14243}{arXiv:2607.14243} (2026).

\bibitem{DeGiorgiPasari2026}
A.~de Giorgi and D.~Pasari,
``The Dark Dimension and Majorana Neutrinos,''
arXiv:2609.10668 [hep-ph] (2026).


\bibitem{AnchordoquiVafa2026}
L.~A.~Anchordoqui and C.~Vafa,
``Cosmology \& Astrophysics of the Dark Dimension with Bulk Majorana Neutrinos,''
\href{https://arxiv.org/abs/2610.15209}{arXiv:2610.15209 [hep-ph]} (2026).


\bibitem{Anchordoqui:2025nmb}
L.~A.~Anchordoqui, I.~Antoniadis and D.~Lust,
Fortsch. Phys. \textbf{73} (2025) no.8, e70015
doi:10.1002/prop.70015
[arXiv:2501.11690 [hep-th]].


\bibitem{AhmedLeontaris2026}
W.~Ahmed and G.~K.~Leontaris,
``Probing Two Dark Dimensions through Primordial Black Holes, Gravitational Waves, and Colliders,''
\href{https://arxiv.org/abs/2606.27992}{arXiv:2606.27992 [gr-qc]} (2026).

\bibitem{Dudas2005}
E.~Dudas, C.~Grojean and S.~K.~Vempati,
``Classical running of neutrino masses from six dimensions,''
\href{https://arxiv.org/abs/hep-ph/0511001}{arXiv:hep-ph/0511001}.

\bibitem{MINOS2019}
P.~Adamson \textit{et al.} (MINOS+ Collaboration),
``Search for sterile neutrinos in MINOS and MINOS+ using a two-detector fit,''
Phys. Rev. Lett. \textbf{122} (2019) 091803,
\href{https://arxiv.org/abs/1710.06488}{arXiv:1710.06488}.

\bibitem{DayaBaySterile2024}
F.~P.~An \textit{et al.} (Daya Bay Collaboration),
``Search for a Sub-eV Sterile Neutrino using Daya Bay's Full Dataset,''
Phys. Rev. Lett. \textbf{133} (2024) 051801,
\href{https://arxiv.org/abs/2404.01687}{arXiv:2404.01687}.

\bibitem{Elacmaz2025}
T.~G.~Ela\c{c}maz, I.~Martinez-Soler and Y.~F.~Perez-Gonzalez,
``Updated Constraints on Large Extra Dimensions from Reactor Antineutrino Experiments,''
\href{https://arxiv.org/abs/2510.12900}{arXiv:2510.12900 [hep-ph]} (2025).

\bibitem{ArkaniHamedSchmaltz2000}
N.~Arkani-Hamed and M.~Schmaltz,
``Hierarchies without symmetries from extra dimensions,''
Phys. Rev. D \textbf{61} (2000) 033005,
\href{https://arxiv.org/abs/hep-ph/9903417}{arXiv:hep-ph/9903417}.

\bibitem{MirabelliSchmaltz2000}
E.~A.~Mirabelli and M.~Schmaltz,
``Yukawa hierarchies from split fermions in extra dimensions,''
Phys. Rev. D \textbf{61} (2000) 113011,
\href{https://arxiv.org/abs/hep-ph/9912265}{arXiv:hep-ph/9912265}.

\bibitem{Girmohanta2021}
S.~Girmohanta, R.~N.~Mohapatra and R.~Shrock,
``Neutrino masses and mixing in models with large extra dimensions and localized fermions,''
Phys. Rev. D \textbf{103} (2021) 015021,
\href{https://arxiv.org/abs/2011.01237}{arXiv:2011.01237}.

\bibitem{NuFit2024}
I.~Esteban, M.~C.~Gonzalez-Garcia, M.~Maltoni, I.~Martinez-Soler, J.~P.~Pinheiro and T.~Schwetz,
``NuFit-6.0: Updated global analysis of three-flavor neutrino oscillations,''
JHEP \textbf{12} (2024) 216,
\href{https://arxiv.org/abs/2410.05380}{arXiv:2410.05380}.

\bibitem{DayaBay2023}
F.~P.~An \textit{et al.} (Daya Bay Collaboration),
``Precision Measurement of Reactor Antineutrino Oscillation at Kilometer-Scale Baselines by Daya Bay,''
Phys. Rev. Lett. \textbf{130} (2023) 161802,
\href{https://arxiv.org/abs/2211.14988}{arXiv:2211.14988}.

\bibitem{Kozynets2024}
T.~Kozynets, P.~Eller, A.~Zander, M.~Ettengruber and D.~J.~Koskinen,
``Constraints on non-unitary neutrino mixing in light of atmospheric and reactor neutrino data,''
\href{https://arxiv.org/abs/2407.20388}{arXiv:2407.20388 [hep-ph]} (2024).

\bibitem{KATRINmass2025}
M.~Aker \textit{et al.} (KATRIN Collaboration),
``Direct neutrino-mass measurement based on 259 days of KATRIN data,''
Science \textbf{388} (2025),
\href{https://arxiv.org/abs/2406.13516}{arXiv:2406.13516}.

\bibitem{KATRINsterile2025}
H.~Acharya \textit{et al.} (KATRIN Collaboration),
``Sterile-neutrino search based on 259 days of KATRIN data,''
Nature \textbf{648} (2025) 70--75,
\href{https://arxiv.org/abs/2503.18667}{arXiv:2503.18667}.

\bibitem{Kleesiek2019}
M.~Kleesiek \textit{et al.},
``Beta-Decay Spectrum, Response Function and Statistical Model for Neutrino Mass Measurements with the KATRIN Experiment,''
Eur. Phys. J. C \textbf{79} (2019) 204,
\href{https://arxiv.org/abs/1806.00369}{arXiv:1806.00369}.

\bibitem{McLaughlinNg2001}
G.~C.~McLaughlin and J.~N.~Ng,
``The Use of nuclear beta decay as a test of bulk neutrinos in extra dimensions,''
Phys. Rev. D \textbf{63} (2001) 053002,
\href{https://arxiv.org/abs/nucl-th/0003023}{arXiv:nucl-th/0003023}.

\bibitem{Basto2013}
V.~S.~Basto-Gonzalez, A.~Esmaili and O.~L.~G.~Peres,
``Kinematical Test of Large Extra Dimension in Beta Decay Experiments,''
Phys. Lett. B \textbf{718} (2013) 1020--1023,
\href{https://arxiv.org/abs/1205.6212}{arXiv:1205.6212}.

\bibitem{TRISTAN2019}
S.~Mertens \textit{et al.} (KATRIN Collaboration),
``A novel detector system for KATRIN to search for keV-scale sterile neutrinos,''
J. Phys. G \textbf{46} (2019) 065203,
\href{https://arxiv.org/abs/1810.06711}{arXiv:1810.06711}.

\bibitem{TRISTAN2026}
H.~Acharya \textit{et al.} (KATRIN Collaboration),
``KATRIN Sensitivity to keV Sterile Neutrinos with the TRISTAN Detector Upgrade,''
\href{https://arxiv.org/abs/2603.23256}{arXiv:2603.23256} (2026).

\bibitem{Bhattacharyya2003}
G.~Bhattacharyya, H.~V.~Klapdor-Kleingrothaus, H.~Pas and A.~Pilaftsis,
``Neutrinoless double beta decay from singlet neutrinos in extra dimensions,''
Phys. Rev. D \textbf{67} (2003) 113001,
\href{https://arxiv.org/abs/hep-ph/0212169}{arXiv:hep-ph/0212169}.

\bibitem{BlennowSeesaw2010}
M.~Blennow, E.~Fernandez-Martinez, J.~Lopez-Pavon and J.~Menendez,
``Neutrinoless double beta decay in seesaw models,''
JHEP \textbf{07} (2010) 096,
\href{https://arxiv.org/abs/1005.3240}{arXiv:1005.3240}.

\bibitem{Rodejohann2011}
W.~Rodejohann,
``Neutrino-less Double Beta Decay and Particle Physics,''
Int. J. Mod. Phys. E \textbf{20} (2011) 1833--1930,
\href{https://arxiv.org/abs/1106.1334}{arXiv:1106.1334}.

\bibitem{Dolinski2019}
M.~J.~Dolinski, A.~W.~P.~Poon and W.~Rodejohann,
``Neutrinoless Double-Beta Decay: Status and Prospects,''
Ann. Rev. Nucl. Part. Sci. \textbf{69} (2019) 219--251,
\href{https://arxiv.org/abs/1902.04097}{arXiv:1902.04097}.

\bibitem{Abada2019}
A.~Abada, A.~Hernandez-Cabezudo and X.~Marcano,
``Beta and Neutrinoless Double Beta Decays with KeV Sterile Fermions,''
JHEP \textbf{01} (2019) 041,
\href{https://arxiv.org/abs/1807.01331}{arXiv:1807.01331}.

\bibitem{Faessler2014}
A.~Faessler, M.~Gonzalez, S.~Kovalenko and F.~Simkovic,
``Arbitrary mass Majorana neutrinos in neutrinoless double beta decay,''
Phys. Rev. D \textbf{90} (2014) 096010,
\href{https://arxiv.org/abs/1408.6077}{arXiv:1408.6077}.

\bibitem{Bolton2025}
P.~D.~Bolton, F.~F.~Deppisch, M.~Rai and Z.~Zhang,
``Probing the nature of Heavy Neutral Leptons in direct searches and neutrinoless double beta decay,''
Nucl. Phys. B \textbf{1010} (2025) 116785,
\href{https://arxiv.org/abs/2212.14690}{arXiv:2212.14690}.

\bibitem{KamLANDZen2025}
S.~Abe \textit{et al.} (KamLAND-Zen Collaboration),
``Search for Majorana Neutrinos with the Complete KamLAND-Zen Dataset,''
Phys. Rev. Lett. \textbf{135} (2025) 262501,
\href{https://arxiv.org/abs/2406.11438}{arXiv:2406.11438}.


\bibitem{Vincent2015}
A.~C.~Vincent, E.~F.~Martinez, P.~Hernandez, M.~Lattanzi and O.~Mena,
``Revisiting cosmological bounds on sterile neutrinos,''
JCAP \textbf{04} (2015) 006,
\href{https://arxiv.org/abs/1408.1956}{arXiv:1408.1956}.

\bibitem{Anchordoqui2025}
L.~A.~Anchordoqui, I.~Antoniadis, D.~Lust and K.~P.~Castillo,
``Cosmological constraints on dark neutrino towers,''
Phys. Rev. D \textbf{111} (2025) 015024,
\href{https://arxiv.org/abs/2411.07029}{arXiv:2411.07029}.


\bibitem{McKeen2024}
D.~McKeen, J.~N.~Ng and M.~Shamma,
``Signatures of bulk neutrinos in the early Universe,''
Phys. Rev. D \textbf{110} (2024) 083507,
\href{https://arxiv.org/abs/2406.05266}{arXiv:2406.05266}.



\bibitem{Abazajian2003}
K.~Abazajian, G.~M.~Fuller and M.~Patel,
``The Cosmological Bulk Neutrino Catastrophe,''
Phys. Rev. Lett. \textbf{90} (2003) 061301,
\href{https://arxiv.org/abs/hep-ph/0011048}{arXiv:hep-ph/0011048}.


\end{thebibliography}
\end{document}